\documentclass[12pt]{article}

\usepackage[letterpaper,hmargin=1.2 in,vmargin=1.0 in]{geometry}
\usepackage{amssymb,amsmath,amsfonts,array,caption,color,comment,eurosym,float, geometry,graphicx,lscape,lipsum,natbib,pdflscape,rotating,sectsty,setspace,subcaption,threeparttable,ulem,url,xargs}
\usepackage[pdftex,dvipsnames]{xcolor}
\usepackage[multiple,hang,flushmargin]{footmisc}
\usepackage{booktabs}
\usepackage{tabularx}
\usepackage{standalone}
\usepackage[font=normalsize,labelfont={bf},font={bf}]{caption}
\usepackage{enumerate}
\usepackage{setspace}

\newcolumntype{L}[1]{>{\raggedright\let\newline\\arraybackslash\hspace{0pt}}m{#1}}
\newcolumntype{C}[1]{>{\centering\let\newline\\arraybackslash\hspace{0pt}}m{#1}}
\newcolumntype{R}[1]{>{\raggedleft\let\newline\\arraybackslash\hspace{0pt}}m{#1}}
\newtheorem{corollary}{Corollary}

\newtheorem{lemma}{Lemma}

\newtheorem{proposition}{Proposition}

\newtheorem{empirical implication}{Empirical Implication}
\newenvironment{proof}[1][Proof]{\noindent\textbf{#1.} }{\ \rule{0.5em}{0.5em}}

\begin{document}

\title{Optimally designing purpose and meaning at work\thanks{
Esther Hauk acknowledges financial support from the Severo Ochoa Programme
for Centers of Excellence in R\&D (Barcelona School of Economics
CEX2024-001476-S), funded by MCIN/AEI/10.13039/501100011033 and from the
Plan Estatal de Investigaci\'{o}n Cient\'{\i}fica, T\'{e}cnica y de Innovaci%
\'{o}n 2024-2027 through project PID2024-156337NB-I00 supported by the
Spanish Ministry of Science, Innovation and Universities / State Research
Agency (MCIN/AEI) and the European Union. }}
\author{Antonio Cabrales\thanks{%
Department of Economics, Universidad Carlos III de Madrid; e-mail:
antonio.cabrales@uc3m.es}, Esther Hauk\thanks{%
Instituto de An\'{a}lisis Econ\'{o}mico (IAE-CSIC) and Barcelona School of
Economics, Campus UAB, 08193 Bellaterra (Barcelona), email:
esther.hauk@iae.csic.es}}
\maketitle

\begin{abstract}
Many workers value purpose and meaning in their jobs alongside income, and
firms need to align these preferences with profit goals. This paper develops
a dynamic model in which firms invest in "purpose" to enhance job meaning
and motivate effort. Workers, who differ in productivity, choose both
productive and socialization effort, gaining utility from income and
meaning. Purpose accumulates over time through firm investment and interacts
with socialization to generate meaning, which boosts productivity. Firms
invest in purpose only insofar as it raises profits. We characterize the
unique equilibrium, including steady state and transition dynamics. Meaning
and purpose rise with the importance workers place on meaning and with
firm's patience, but fall with depreciation and socialization costs. The
relationship with workers' share of output is non-monotonic. We also show
that some intermediate level of heterogeneity in skills is best for
performance. Compared to a worker-owned firm, profit-maximizing firms
underinvest in purpose, highlighting a misalignment between firm incentives
and worker preferences. The model provides insight into when and why firms
adopt purpose-driven practices and underscores the role of diversity in
fostering meaning at work.

{\small \noindent \textbf{JEL Classification}: M50, M52, L23.\newline
}
\end{abstract}

\section{Introduction}

There is a large body of evidence suggesting that individuals seek more from
their jobs than financial compensation alone. Most workers aspire to roles
that also offer a sense of purpose and meaning. As Blount and Leinwand
(2019) argue, \textquotedblleft Over the past decade, \textquotedblleft
purpose\textquotedblright\ has become a management watchword. Since 2010 it
has appeared in the titles of more than 400 new business and leadership
books and thousands of articles.\textquotedblright\ Despite this increased
attention, aligning the search for purpose with a profit-making objective
remains challenging. The same authors observe that a global survey of over
540 employees conducted by Strategy\& found that only 28\% felt fully
connected to their company's stated purpose. This disconnect highlights the
difficulty firms face in incorporating meaning into the workplace in a way
that aligns with profitability.

Firms have experimented with various strategies to bridge this gap. Some
allow employees to allocate part of their time to self-directed projects
that foster ownership and creativity, while others encourage engagement in
community or charitable work during paid hours. In those activities,
employees often collaborate with others in the company. Companies like
Microsoft, Google, Salesforce, and Timberland offer structured programs in
this spirit, sometimes matching employees' time donations with monetary
contributions and promoting team-based participation. Empirical evidence
suggests that these efforts are not merely symbolic: they are associated
with gains in productivity and profitability (Veleva et al. 2012; Bo\v{s}tjan%
\v{c}i\v{c} et al. 2018; Knox 2020).

This paper develops a formal model to understand how firms make optimal
decisions about investing in purpose, taking into account that workers value
meaning and adjust their effort accordingly. The model is inspired by the
frameworks of Cabrales, Calv\'{o}-Armengol and Zenou (2011) and Albornoz,
Cabrales and Hauk (2019), and introduces two forms of worker effort:
productive effort and socialization effort. Both are costly, but both
contribute to firm output. In addition, firms can make costly investments in
\textquotedblleft purpose,\textquotedblright\ which accumulates over time to
generate an increased stock of job \textquotedblleft
meaning\textquotedblright\ and interacts with workers' socialization
efforts. Meaning, in turn, enhances worker utility and productivity, further
incentivizing effort. That is, we call purpose the flow that converts into
meaning, which is the stock that enhances utility and productivity.

We consider a setting where firms are infinitely lived and workers live for
one period. Firms and workers make optimal decisions simultaneously within
each period. Workers receive a share of output, derive utility from both
income and job meaning, and bear the costs of effort. Firms maximize
profits, defined as output net of worker compensation and investment in
purpose. Although firms do not value meaning per se, they invest in it when
it contributes to increased productivity and, consequently, profitability.
Worker productivity is heterogeneous, follows a continuous distribution, and
is complementary with both effort and meaning.

We fully characterize the unique equilibrium of the model, including the
steady state and the transition dynamics. Comparative statics reveal several
key insights. In steady state, both job meaning and firm investment in
purpose increase with the importance workers place on meaning in their
utility function and with the firm's discount factor, and decline with the
depreciation rate of meaning. Remember that meaning is the stock that
enhances value, and it increases by investing in purpose. Hence the
importance of the discount factor and the depreciation rate, in both the
investment in purpose (the flow) and the final accumulated value (the
meaning). It is also not surprising that when workers value meaning more,
both the stock and the flow increase. These variables also decrease with the
marginal cost to firms of providing purpose and with the cost of
socialization effort to workers, but are unaffected by the marginal cost of
productive effort. To understand this, notice that meaning is complementary
to socialization effort but not to productive effort. The relationship of
meaning and purpose with the worker's share of output is non-monotonic: both
meaning and purpose rise with the share, provided that the overall weight of
income and meaning in worker utility does not exceed a critical threshold.
The reason for this is that an increase in worker's share increases the
socialization effort directly, and through that, profitability. However,
when the worker's share becomes too large, firm profits decline
significantly, reducing investment in purpose. This, in turn, lowers
socialization efforts and diminishes meaning.

To characterize the comparative statics of the ability variable, we consider
the effects of a first order and second order stochastic dominant shift in
the distribution of that variable. While a first order stochastic dominance
shift always increases purpose and meaning, the effects of a second order
stochastic dominance shift are nuanced. We show that increasing dispersion
increases meaning, purpose, individual utility, and firm profits when the
initial distribution is not too dispersed, but decreases them when the
initial distribution is highly dispersed. We can conclude that some
intermediate level of heterogeneity in skills seems to be best for
performance. These results highlight the importance of diversity in ability
for fostering purpose in the workplace. Notice that the diversity in our
model is such that, everything else equal, some workers are more productive
than others. In a sense, we are talking about \emph{vertical}
differentiation. Due to the interaction between output and the talent
distribution in our model, these effects can be complex.

The model also yields implications for firm profits and worker welfare when
other parameters change. Profits and worker welfare increase with the value
workers place on meaning and the firm's discount factor, and decrease with
the depreciation rate of meaning, the firm's cost of providing purpose, and
workers' effort costs. To understand this, notice that the utility of the
worker and firm profits are increasing in the equilibrium level of purpose,
meaning, and the firm production, which depend on production and
socialization efforts. So these comparative statics are derived from those
relating to investment in purpose, meaning, and the socialization efforts,
discussed above.

The effect of the worker's output share on profits and worker welfare is
ambiguous, but they increase with the share provided the combined weight on
income and meaning in utility remains moderate (although this threshold
combined weight would be different for profits and worker utility). This
condition ensures that even a fully profit-maximizing firm would choose to
allocate a positive share of output to workers. These results are related to
the fact that the worker share increases socialization effort directly,
incentivizing investments in purpose and through it increasing the stock of
meaning. But when the worker share is too high, there is a reduction in
profits reducing investment, which also hurts the worker utility.

Finally, we compare the benchmark case of profit-maximizing firms with a
fully worker-owned firm. While the former does invest in purpose to induce
effort, we find that workers would prefer a higher level of investment in
meaning. This gap reflects a fundamental divergence between firm incentives
and worker preferences in the provision of purpose, and suggests that
ownership structure may play a significant role in shaping the extent to
which firms cultivate meaningful work environments.

The rest of the paper is structured as follows: the next Section discusses
the related literature. Section 3 presents and solves the model. In Sections
4 and 5 we perform the comparative static analysis. Section 6 analyzes the
path to steady state. In Section 7 we solve for the worker-owned firm.
Section 8 concludes and discusses future research.

\section{Literature}

This paper contributes to a growing literature on the role of purpose and
meaning in the workplace, particularly in relation to worker motivation and
firm performance. Much of the economic theory literature has focused on how
organizations sort or match individuals with different intrinsic motivations
to firm missions. For example, Besley and Ghatak (2005) develop a model of
moral hazard in which both principals and agents have mission preferences,
and organizational alignment reduces agency frictions. Delfgaauw and Dur
(2007) similarly consider a model with unobservable worker types, where
contracts are used to screen for intrinsic motivation. Delfgaauw and Dur
(2008) extend this framework to account for differences between public and
private sector employment. In Cassar and Armouti-Hansen (2020) firms use
meaning strategically to attract and motivate the "right" types of workers.

Our approach departs from this class of models in two important ways. First,
we do not focus on sorting or matching between heterogeneous worker and firm
types. Instead, we assume all workers value meaning and ask how
profit-maximizing firms optimally choose investments in purpose. Second, the
existing models are static and treat meaning as a purely individual trait or
alignment. In contrast, we model meaning as a dynamic, accumulable construct
that interacts with worker behavior---specifically socialization effort---to
affect productivity. This focus on the endogenous creation of purpose within
firms is aligned with insights from the management literature, but has
received less attention in economic theory.

Cassar and Meier (2018) offer a theoretical framework in which work meaning
enters utility additively alongside consumption and effort costs, and
decompose meaning into four components: mission, autonomy, competence, and
relatedness. Our model builds on this work by focusing on one empirically
salient component---relatedness---which appears to be the dominant channel
through which meaning affects outcomes. We provide a more structural account
of how relatedness arises through social interaction, how it affects
productivity, and how firms can influence it through investment in purpose.
Our model thus complements and extends Cassar and Meier's (2018) framework
by offering a dynamic, micro-founded treatment of one key dimension of
meaning.

Several recent applications build on Cassar and Meier's (2018) framework.
For instance, Armouti-Hansen et al. (2024) test the model experimentally,
and Kesternich et al. (2021) extend it to incorporate fairness concerns and
work norms. Nikolova and Cnossen (2020) suggest integrating insights from
self-determination theory into this framework, and Cnossen and Nikolova
(2024) develop such a model formally. We share with these papers an emphasis
on non-monetary drivers of worker behavior but differ in our focus on
firm-side dynamics and the strategic provision of meaning over time.

Cassar and Meier (2018) also compile an extensive set of stylized facts from
the empirical literature, which inform the assumptions and mechanisms in our
model. Nikolova and Cnossen (2020), using data from three waves of the
European Working Conditions Survey, show that autonomy, competence, and
relatedness explain around 60\% of the variation in perceived work
meaningfulness. They also construct an index of meaningful work based on
these dimensions. Our model is consistent with these empirical patterns and
gives theoretical substance to the role of relatedness in particular.

Recent empirical work also supports the idea that increasing perceived
meaning enhances performance. Ashraf et al. (2025) conduct a large-scale
randomized controlled trial (RCT) in a multinational firm, where a treatment
group is exposed to an intervention designed to help workers perceive the
meaning of their work. The result is a significant increase in performance
(in relative or within-group terms rather than absolute differences), due
both to improved productivity among low performers and the voluntary exit of
the least motivated employees. These effects are consistent with our model:
we predict higher effort when meaning is more salient and, if a
participation margin were added, our model would likely predict the
departure of low-motivation workers, leading to higher average productivity.

The human resource management literature has long emphasized the importance
of meaning at work, both for individual well-being and organizational
performance. Steger (2016) provides a comprehensive overview of this
literature and emphasizes that meaning is not merely a sorting mechanism but
something that can be cultivated within organizations. This insight is
central to our approach: unlike much of the economics literature, we model
purpose as a strategic variable under firm control, shaped by investment and
by organizational structure.

A growing empirical literature explores the relationship between firm-level
purpose and performance. Gartenberg et al. (2019), using a dataset of nearly
half a million workers, show that firms with a clearly articulated and
widely shared sense of purpose outperform others in terms of stock market
valuation, controlling for current fundamentals. Similarly, Gartenberg
(2023), using a larger sample, finds that firms with high levels of
intangible capital, innovation, and long-term investors are more likely to
pursue purpose alongside profits. These firms do not treat purpose and
profit as competing goals, but rather as complementary. Our model is
consistent with these findings: we posit long-lived, profit-maximizing firms
that invest in purpose because it increases productivity and ultimately
improves long-run returns.

Finally, George et al. (2023) and Jasinenko and Steuber (2023) address
concerns about the vague and often inconsistent definitions of purpose and
meaning in the management literature. They propose validated constructs for
empirical measurement. Our theoretical framework complements this effort by
offering a precise and operational definition of purpose, focused on social
interaction and its productivity-enhancing effects. We emphasize the
relatedness dimension of meaning, in line with empirical evidence, and show
how it interacts with firm behavior and output in a dynamic setting.

\section{The model}

We study how long-lived, profit-maximizing firms choose the optimal level of
purpose in an environment where creating purpose entails costs but enhances
worker motivation and, consequently, output. In our framework, purpose is
not pursued for its own sake but as a strategic investment that complements
traditional incentive mechanisms. The model reflects growing empirical
evidence that purpose and profitability are not inherently at odds.
Gartenberg et al. (2019) show that a stronger sense of purpose is associated
with superior financial performance, while Gartenberg (2023) finds that this
relationship is particularly pronounced in firms with long-term
owners---precisely the type of firms our model is designed to capture. In
line with these findings, we formalize how investment decisions on purpose
can arise endogenously as part of a firm's optimal profit-maximizing
strategy.

The agents in our model are a firm and its workers. The firm maximizes its
long-run profit by choosing how much to invest in its mission/purpose $r_{t}$
each period $t$. The cost of providing purpose is given by 
\begin{equation*}
C\left( r_{t}\right) =\frac{C}{2}r_{t}^{2}.
\end{equation*}%
The variable $r_{t}$ is chosen by the firm at the beginning of the period
and is observed by the workers at the time they make their choices.

Workers only live one period and each worker $i$ maximizes their utility
choosing how much work effort $e_{i,t}$ to put into their tasks and how much
socialization effort $k_{i,t}$ they put into interacting with their
co-workers. Both types of effort are costly and the cost functions are given
by%
\begin{equation*}
c_{i,t}^{e}=\frac{A_{e}}{2}e_{i,t}^{2}\text{ and }c_{i,t}^{k}=\frac{A_{k}}{2}%
k_{i,t}^{2}
\end{equation*}%
where $A_{e}$ and $A_{k}$ are positive constants.

Work effort generates a productive return. In production, the socialization
input $s_{i,t}$ stems from socialization efforts and is defined as: 
\begin{equation}
s_{i,t}=k_{i,t}^{1/2}\int k_{j,t}^{1/2}d_{j}  \label{socialization}
\end{equation}%
This input boosts productivity by strengthening the sense of meaning at
work. Its specific functional form can be microfounded as in Albornoz et al.
(2019), by assuming symmetry, anonymity and constant returns in
socialization. The degree to which socialization fosters meaning depends on
the firm's chosen level of purpose $r_{t}$. Such investment may involve
initiatives like granting employees time to pursue personal projects or
engage in volunteer activities. As documented by e.g. Veleva et al. (2012),
Bo\v{s}tjan\v{c}i\v{c} et al. (2018), Knox (2020), this possibility has a
positive impact on both productivity and motivation.

The purpose variable $r_{t}$ is the flow of resources that forms the basis
of the total job meaning of worker $i$ in period $t$, $m_{i,t}$ which we
assume is given by\footnote{$r_{t}$ has decreasing returns in meaning. This
is probably realistic, but also technically important. Since $k_{i,t}$ will
be influenced by $r_{t}$, the overall influence of $r_{t}$ on output
exhibits weaker diminishing returns. If these returns were substantially
higher, they would undermine the concavity of the dynamic programming
problem.
\par
{}}

\begin{equation*}
m_{i,t}=s_{i,t}r_{t}^{\frac{1}{2}}+\lambda \overline{m}_{t-1},
\end{equation*}%
which can be rewritten using (\ref{socialization}) as 
\begin{equation}
m_{i,t}=k_{i,t}^{1/2}\int k_{j,t}^{1/2}d_{j}r_{t}^{\frac{1}{2}}+\lambda 
\overline{m}_{t-1}  \label{individual meaning}
\end{equation}
Note that that meaning in period $t$ for individual $i,$ is influenced by
average meaning in the group, from the previous period $\overline{m}_{t-1}$ 
\begin{equation*}
\overline{m}_{t-1}=\int m_{j,t-1}d_{j}
\end{equation*}%
with a persistency factor $\lambda $. A higher $\lambda $ implies that
meaning decays (depreciates) more slowly from one period to the next. The
concept of decay may be interpreted through the lens of employee turnover.
The persistence of organizational culture is fundamentally contingent upon
the tenure of employees, as sustained presence is required for cultural
investments to become internalized within individual dispositions and
operational practices. Firms characterized by high turnover and short tenure
are therefore likely to exhibit a greater rate of cultural decay.

A central feature of our model is that job meaning arises through
socialization efforts, rather than through directly productive efforts. This
distinction is grounded in the human resources literature, which
consistently emphasizes the social and relational dimensions of meaningful
work. As Steger (2016) notes, individuals experience greater meaning when
they transcend self-oriented concerns and engage with others in prosocial
ways (see also Dik, Duffy, \& Steger, 2012; Reker, Peacock, \& Wong, 1987;
Schnell, 2009; Steger, Kashdan, \& Oishi, 2008). In his framework, the
highest level of meaningful work is associated with contributions to others
or to the greater good. Furthermore, Steger (2016) emphasizes that, at the
interpersonal level, meaningful work is fostered by respectful
relationships, understanding of the organizational social context, and
opportunities for mutual support, such as helping, mentoring, or being
mentored. Our modeling choice reflects these insights by linking meaning to
social interaction within the firm, rather than to the execution of
individual productive tasks.

Meaning also exerts an independent influence on worker utility, as employees
tend to derive greater satisfaction from tasks they perceive as meaningful.
This reflects that the value of such work extends beyond its immediate
output, encompassing broader personal and social impacts. A growing body of
empirical research supports the relevance of this modeling choice. For
example, Grueso et al. (2022), in their study of a supported employment
program in Colombia, find a strong positive correlation between perceived
meaningfulness of work and overall life satisfaction. Similarly, Allan et
al. (2018) report that meaningful work is associated with lower levels of
depression, particularly when accompanied by job satisfaction. Additional
evidence indicates that meaningful work contributes to reduced stress and
anxiety. Wandycz-Mej\'{\i}as et al. (2025) further show that employees who
perceive their work as meaningful are less likely to express intentions to
leave their organization and more likely to report high levels of job
satisfaction. As Steger (2016) aptly notes: \textquotedblleft If the primary
incentive a company is offering to a worker to work hard and stay with the
company is monetary, then what is to prevent the worker from taking a better
deal elsewhere?\textquotedblright\ This body of evidence underscores the
importance of intrinsic motivation---particularly through meaningful
work---in shaping employee behavior and retention, thereby justifying its
inclusion in the worker's utility function.

Denoting the output of the worker $i$ by $z_{i,t}\left(
m_{i,t},e_{i,t}\right) $, and the worker $i$'s ability by $b_{i,t}$ (which
are distributed according to some distribution function), the utility of the
worker $i$ in period $t$ can be written as:%
\begin{equation}
u_{i,t}(e_{i,t},k_{i,t})=\beta b_{i,t}m_{i,t}+\alpha z_{i,t}\left(
m_{i,t},e_{i,t}\right) -\frac{A_{e}}{2}e_{i,t}^{2}-\frac{A_{k}}{2}k_{i,t}^{2}
\label{Utility worker}
\end{equation}%
where the term $\beta b_{i,t}m_{i,t}$, with $\beta >0,$ captures the
intrinsic utility that a worker derives from job meaning, independent of
material compensation. The parameter $\alpha $ denotes the share of output
that the worker is entitled to retain, thereby determining the worker's
compensation. This share reflects the worker's bargaining power within the
framework of Nash bargaining, which governs the negotiation process between
firms and workers, after output has been produced. The costs of $r_{t}$ and
efforts, $c_{i,t}^{e}$ and $c_{i,t}^{k}$ are not included in the negotiation
as they are not contractible.

From now on we suppress the integration variable for ease of notation.

Output produced by worker $i$ in period $t$ is 
\begin{eqnarray}
z_{i,t}\left( m_{i,t},e_{i,t}\right) &=&b_{i,t}\left(
e_{i,t}\,+m_{i,t}\right)  \notag \\
&=&b_{i,t}\left( e_{i,t}\,+k_{i,t}^{1/2}\int k_{j,t}^{1/2}r_{t}^{\frac{1}{2}%
}+\lambda \overline{m}_{t-1}\right)  \label{worker output}
\end{eqnarray}%
Notice that effort and meaning are perfect substitutes in this setup and are
complements to ability $b_{i,t}.$

Thus, every period starts with the realized average meaning from last
period, $\overline{m}_{t-1}$. Then, the firm chooses purpose $r_{t}$ for the
period. After that, the workers choose $e_{i,t}\,$and $k_{i,t}.$Then output $%
z_{i,t}$ and payoffs realize, and $\overline{m}_{t}$ is generated.

\subsection{The decision of worker $i$}

Introducing (\ref{individual meaning}), (\ref{socialization}) and (\ref%
{worker output}) into (\ref{Utility worker}), the worker optimization
problem is to choose $e_{i,t}$ and $k_{i,t}$ to maximize 
\begin{equation}
u_{i,t}(e_{i,t},k_{i,t})=\left( \alpha +\beta \right) b_{i,t}\left( \left(
k_{i,t}^{1/2}\int k_{j,t}^{1/2}\right) r_{t}^{\frac{1}{2}}+\lambda \overline{%
m}_{t-1}\,\right) +\alpha b_{i,t}e_{i,t}-\frac{A_{e}}{2}e_{i,t}^{2}-\frac{%
A_{k}}{2}k_{i,t}^{2}  \label{utility worker optimize}
\end{equation}

The resulting first order conditions are: 
\begin{equation}
\alpha b_{i,t}-A_{e}e_{i,t}=0\text{ for }e_{i,t}  \label{FOCe}
\end{equation}%
\begin{equation}
\frac{1}{2}\left( \alpha +\beta \right) b_{i,t}\left( r_{t}^{\frac{1}{2}%
}\left( k_{i,t}^{-1/2}\int k_{j,t}^{1/2}\right) \right) -A_{k}k_{i,t}=0\text{
for }k_{i,t}\text{. }  \label{FOCk}
\end{equation}

\begin{lemma}
\label{lemma worker problem}Each worker $i$'s choice problem has a unique
(interior) solution. In particular 
\begin{eqnarray}
e_{i,t} &=&b_{i,t}e_{t}\text{ where }e_{t}=\frac{\alpha }{A_{e}}
\label{work effort} \\
k_{i,t} &=&b_{i,t}^{\frac{2}{3}}k_{t}\text{ where }k_{t}=\frac{\alpha +\beta 
}{2A_{k}}r_{t}^{\frac{1}{2}}\left( \int b_{j}^{\frac{1}{3}}\right)
\label{socialization effort}
\end{eqnarray}%
where $e_{t}$ and $k_{t}$ are identical to all workers of the firm.
\end{lemma}

\begin{proof}
See the Appendix.
\end{proof}

\begin{empirical implication}
Socialization effort is complementary to the firm's investment in purpose,
i.e., when purpose increases, so does socialization effort.
\end{empirical implication}

This is important to understand why factors that lead to an increase of
investment in purpose will also increase socialization effort and vice versa.

Allan et al. (2019) show in a meta-analysis of 44 papers that meaningful
work correlates positively with outcomes like organizational citizenship
behavior (OCBs) (i.e., discretionary helpful, cooperative behavior toward
coworkers).

Lemma \ref{lemma worker problem} also implies that more talented individuals
make higher productive and socialization efforts. In addition, the
individual talent in the group has externalities, since the average talent
in the group also increases socialization effort.

\begin{empirical implication}
Average talent is complementary to the firm's investment in purpose, which
is more impactful in highly talented groups.
\end{empirical implication}

Gartenberg et al. (2019) show that meaningfulness positively affects
performance where employees have discretion and complex problem-solving in
their roles (mid-level/professional jobs).

Having solved each worker's maximization problem we are now in a position to
rewrite the meaning for each worker $i$ in his job in period $t$ (\ref%
{individual meaning}) as a function of $r_{t}$

\begin{equation*}
m_{i,t}=k_{i,t}^{1/2}\int k_{j,t}^{1/2}r_{t}^{\frac{1}{2}}+\lambda \overline{%
m}_{t-1}=\frac{\alpha +\beta }{2A_{k}}r_{t}\left( b_{i,t}^{\frac{1}{3}%
}\right) \left( \int b_{j,t}^{\frac{1}{3}}\right) ^{2}+\lambda \overline{m}%
_{t-1}
\end{equation*}

\subsection{Decision of the firm in terms of $r_{t}$}

While the workers maximize their one period utility, the firm chooses $r_{t}$
to maximize its lifetime utility given by:

\begin{equation}
\sum_{t,i\,_{t}}\delta ^{t-1}\left( 1-\alpha \right) \left( z_{i,t}\left(
m_{i,t},e_{i,t}\right) \right) -C\left( r_{t}\right)  \label{life-time firm}
\end{equation}%
Using the worker's optimal socialization and work efforts, each worker's
output can be expressed as

\begin{equation}
z_{i,t}\left( m_{i,t},e_{i,t}\right) =b_{i,t}\left( b_{i,t}\frac{\alpha }{%
A_{e}}+\frac{\alpha +\beta }{2A_{k}}r_{t}\left( b_{i,t}^{\frac{1}{3}}\right)
\left( \int b_{j,t}^{\frac{1}{3}}\right) ^{2}+\lambda \overline{m}%
_{t-1}\right)  \label{individual output}
\end{equation}

Notice that in equation (\ref{individual output}), the individual talent
parameter $b_{i,t}$ appears squared in the first term and raised to the
power of $4/3$ in the second. When this is aggregated to the total output of
the firm, we will have terms with $\int b_{j,t}^{\frac{4}{3}} $ and $\int
b_{j,t}^{2},$ in addition to $\int b_{j,t}^{\frac{1}{3}}$ which comes from $%
k_{t}$.

\begin{empirical implication}
More talented groups are more productive, and past meaning ($\overline{m}%
_{t-1}$) is particularly important for the more productive individuals.
\end{empirical implication}

This suggests that young talented workers might be especially attracted to
high meaning firms. In two online-labor-market field experiments Burbano
(2016) found that telling applicants the employer was socially responsible
reduced the wage premium demanded by the highest performers---i.e., top
performers were willing to work for less when the job signaled
meaning/purpose.

As we have seen, talented people make higher efforts, and they also create
externalities in the group. In addition, the fact that different moments of
the distribution of $b_{j,t}$ affect the output will mean that the impact of
changes in its dispersion (which we can call diversity) are also relevant,
in nuanced ways we will discuss later.

Introducing individual productivity (\ref{individual output}) into (\ref%
{life-time firm}), the firm's optimization problem is to choose $\alpha $
and $r_{t}$ each period to maximize 
{\footnotesize \begin{equation}
\sum_{t}\left( \delta ^{t-1}\left( 1-\alpha \right) \left( \int b_{j,t}^{2}%
\frac{\alpha }{A_{e}}+\frac{\alpha +\beta }{2A_{k}}r_{t}\left( \int b_{j,t}^{%
\frac{4}{3}}\right) \left( \int b_{j,t}^{\frac{1}{3}}\right) ^{2}+\lambda
\int b_{j,t}\overline{m}_{t-1}\right) \right) {\small -\delta }^{t-1}\frac{C%
}{2}{\small r}_{t}^{2}  \label{firm objective}
\end{equation}
}
where the law of motion of $\overline{m}_{t}$ is given by 
\begin{equation}
\overline{m}_{t}=\frac{\alpha +\beta }{2A_{k}}r_{t}\left( \int b_{j,t}^{%
\frac{1}{3}}\right) ^{3}+\lambda \overline{m}_{t-1}.  \label{law of motion}
\end{equation}

\begin{proposition}
\label{steady state values} In steady state 
\begin{equation}
\overline{m}^{\ast }=\frac{\left( 1-\alpha \right) \int b_{j}^{\frac{4}{3}%
}\left( \alpha +\beta \right) ^{2}\left( \int b_{j}^{\frac{1}{3}}\right) ^{5}%
}{C4A_{k}^{2}\left( 1-\lambda \right) }+\frac{\left( 1-\alpha \right) \delta
\lambda \int b_{j}\left( \alpha +\beta \right) ^{2}\left( \int b_{j}^{\frac{1%
}{3}}\right) ^{6}}{C\left( 1-\delta \lambda \right) 4A_{k}^{2}\left(
1-\lambda \right) }  \label{steady state m}
\end{equation}%
while 
\begin{equation}
r^{\ast }=\frac{\left( 1-\alpha \right) \int b_{j,t}^{\frac{4}{3}}\left(
\alpha +\beta \right) \left( \int b_{j,t}^{\frac{1}{3}}\right) ^{2}}{2CA_{k}}%
+\frac{\left( 1-\alpha \right) \delta \lambda \int b_{j,t}\left( \alpha
+\beta \right) \left( \int b_{j,t}^{\frac{1}{3}}\right) ^{3}}{2CA_{k}\left(
1-\delta \lambda \right) }  \label{steady state r}
\end{equation}%
and 
\begin{equation}
k^{\ast }=\frac{\left( 1-\alpha \right) ^{\frac{1}{2}}\left( \alpha +\beta
\right) ^{\frac{3}{2}}}{2A_{k}}\left( \int b_{j}^{\frac{1}{3}}\right) \left(
\int b_{j}^{\frac{2}{3}}\right) \left( \frac{\int b_{j}^{\frac{4}{3}}\left(
\int b_{j}^{\frac{1}{3}}\right) ^{2}}{2CA_{k}}+\frac{\delta \lambda \int
b_{j}\left( \int b_{j}^{\frac{1}{3}}\right) ^{3}}{2CA_{k}\left( 1-\delta
\lambda \right) }\right) ^{\frac{1}{2}}  \label{steady state k}
\end{equation}
\end{proposition}

\begin{proof}
\bigskip See Appendix
\end{proof}

\section{Comparative statics}

We are now in a position to determine how steady state average meaning in
the firm $\overline{m}^{\ast }$ given by (\ref{steady state m}) as well as
the steady state investment in purpose $r^{\ast }$ given by (\ref{steady
state r}) and the average socialization effort $k^{\ast }$ given by (\ref%
{steady state k}) varies with the underlying parameters.

\begin{proposition}
\label{comparative statics meaning and k}$\overline{m}^{\ast }$, $r^{\ast }$
and $k^{\ast }$ are increasing in $\beta $, in the discount factor $\delta $
and in the persistency factor of meaning $\lambda .$ They are also
decreasing in the firm's marginal cost of providing purpose $C\,$, the
individual marginal socialization cost $A_{k}$ and independent of the
individuals marginal work effort cost $A_{e}$. Concerning the share of
output $\alpha $ given to the workers 
\begin{equation*}
\frac{\partial \overline{m}^{\ast }}{\partial \alpha }>0\Leftrightarrow
2>3\alpha +\beta
\end{equation*}%
while 
\begin{equation*}
\frac{\partial r^{\ast }}{\partial \alpha }>0\Leftrightarrow 1>2\alpha +\beta
\end{equation*}%
and%
\begin{equation*}
\frac{\partial k^{\ast }}{\partial \alpha }>0\Leftrightarrow 3>4\alpha +\beta
\end{equation*}
\end{proposition}

\begin{proof}
See the appendix
\end{proof}

The impact of the discount factor $\delta $ and in the persistency factor of
meaning $\lambda $ on the investment $r^{\ast }$ and the stock $\overline{m}%
^{\ast }$ are easy to understand. Since the socialization effort $k^{\ast }$%
which directly enhances production is complementary to $\overline{m}^{\ast }$
(through $r^{\ast }$), those variables are an investment into the
profitability of the firm. So an improvement in the discount factor or the
persistency factor of meaning will tend to increase them, and through the
complementarity, they also increase $k^{\ast }$. The influence of $\alpha $
is naturally more nuanced. On the one hand, we can see from Lemma \ref{lemma
worker problem} that an increase in $\alpha $ increases $k^{\ast }$directly,
and through that, profitability. However, when $\alpha $ becomes excessively
large, firm profits decline substantially, which in turn constrains
investment $r^{\ast }$. This reduction subsequently leads to decreases in $%
k^{\ast }$ and $\overline{m}^{\ast }$.

Proposition \ref{comparative statics meaning and k} has a series of
empirical implications:

\begin{empirical implication}
Firms who emphasize the long-run horizon invest more in purpose.
\end{empirical implication}

Empirical evidence links a long-term orientation to greater investment in
purpose-driven, meaning-enhancing practices. Using a regression
discontinuity on close-call shareholder votes for long-term executive
compensation, Flammer and Bansal (2017) show that firms pushed toward longer
horizons subsequently increase stakeholder engagement - especially toward
employees and the natural environment - and see operating gains two years
later, consistent with purposeful, non-myopic investment patterns. At the
cultural level, a large-sample study of Chinese listed firms by Fang, Wen,
and Xu (2024) text-mines annual reports to construct a long-term orientation
culture index and finds that more long-term oriented firms boost innovation
via mechanisms such as upgrading employees' educational qualifications and
strengthening internal controls, with effects intensifying under performance
pressure - signals of organization-wide investments that support meaningful
work systems.

\begin{empirical implication}
Investment in meaning is decreasing with the employee turnover in a sector.
\end{empirical implication}

Sectors with higher employee turnover tend to invest less in initiatives
that enhance job meaning or purpose, as the expected return on such
investments is discounted when employee tenure is short. Retail,
hospitality, and call centers in the US exemplify the pattern. Retail and
hospitality have historically relied on extrinsic motivators such as pay
adjustments and scheduling flexibility, while purpose-driven strategies and
intrinsic motivators like autonomy and development opportunities remain
limited (Gitnux, 2025; Work Institute, 2025). Call centers, where annual
turnover averages 40--45\% and first-year attrition can exceed 70\%, face
similar challenges; although engagement programs exist, they often emphasize
operational efficiency rather than intrinsic motivators, despite evidence
that autonomy, mastery, and purpose significantly improve retention
(Insignia Resources, 2025; Anwar et al., 2018, Farrell, 2024). In contrast,
high-tech industries - where voluntary turnover averages around 13\%, far
below these sectors - invest heavily in intrinsic motivators such as
meaningful projects, autonomy, and career development, recognizing their
role in innovation and retention (Mercer, 2025, Vazquez, 2025).

\begin{empirical implication}
There is an inverted U-shaped / relationship between workers' share of
output and their level of socialization, as well as with the meaning and
purpose.
\end{empirical implication}

We have not been able to find studies that are directly testing this
implication, but there is some evidence pointing in this direction. Wang et
al. (2015) show that pay dispersion, which is somewhat related to the
workers' share in revenues, and employee participation show an inverted-U
connection. In firm data, moderate within-firm pay gaps maximize
participation, which in turn boosts innovation - both are
socialization/cooperation outcomes. This implication is intriguing and it
may be a relevant target of future research.

In the next subsection we show that these comparative statics are largely
inherited in firm profits and worker utility, confirming the above intuition.

\subsection{Worker utility and firm profits in steady state}

Introducing (\ref{work effort}) and (\ref{socialization effort}), the worker
utility (\ref{utility worker optimize}) in steady state can be rewritten as:%
\begin{eqnarray}
u^{\ast } &=&b_{i,t}^{\frac{4}{3}}\left( \frac{\left( \alpha +\beta \right)
^{2}}{2A_{k}}\left( \int b_{j}^{\frac{1}{3}}\right) ^{2}\right) r^{\ast
}+\lambda (\alpha +\beta )b_{i,t}\overline{m}^{\ast }\,  \notag \\
&&+\alpha b_{i,t}^{2}\frac{\alpha }{A_{e}}-\frac{A_{e}}{2}b_{i,t}^{2}\left( 
\frac{\alpha }{A_{e}}\right) ^{2}-\frac{A_{k}}{2}b_{i,t}^{\frac{4}{3}}\left( 
\frac{\alpha +\beta }{2A_{k}}\right) ^{2}r^{\ast }\left( \int b_{j}^{\frac{1%
}{3}}\right) ^{2}  \notag \\
&=&b_{i,t}^{\frac{4}{3}}\left( \frac{3\left( \alpha +\beta \right) ^{2}}{%
8A_{k}}\left( \int b_{j}^{\frac{1}{3}}\right) ^{2}\right) r^{\ast }+\lambda
(\alpha +\beta )b_{i,t}\overline{m}^{\ast }\,+\frac{b_{i,t}^{2}}{2}\frac{%
\alpha ^{2}}{A_{e}}  \label{steady state utility}
\end{eqnarray}

From Proposition \ref{comparative statics meaning and k} it is easy to see
that steady state worker utility is increasing in $\beta $, in the discount
factor $\delta $ and in the persistency factor of meaning $\lambda $. It is
decreasing in the firm's marginal cost of providing purpose $C\,$, the
individual marginal socialization cost $A_{k}$ and decreasing in the
individuals marginal work effort cost $A_{e}$. These results are easy to
understand by noticing that the utility of the worker is increasing in $%
\overline{m}^{\ast }$, $r^{\ast }$ and the firm production, which depends on
production and socialization efforts. So these comparative statics are
inherited from those of $\overline{m}^{\ast }$, $r^{\ast }$ and $k^{\ast }$
that are discussed in Proposition \ref{comparative statics meaning and k}.

Concerning the share of output $\alpha $ given to the workers

\begin{eqnarray*}
\frac{\partial u^{\ast }}{\partial \alpha } &=&b_{i,t}^{\frac{4}{3}}\left(
\left( \frac{3\left( \alpha +\beta \right) }{4A_{k}}\left( \int b_{j}^{\frac{%
1}{3}}\right) ^{2}\right) r^{\ast }+\left( \frac{3\left( \alpha +\beta
\right) ^{2}}{8A_{k}}\left( \int b_{j}^{\frac{1}{3}}\right) ^{2}\right) 
\frac{\partial \overline{r}^{\ast }}{\partial \alpha }\,\right) \\
&&+\lambda b_{i,t}\overline{m}^{\ast }+\lambda (\alpha +\beta )b_{i,t}\frac{%
\partial \overline{m}^{\ast }}{\partial \alpha }+b_{i,t}^{2}\frac{\alpha }{%
A_{e}}
\end{eqnarray*}%
a sufficient condition $\frac{\partial u^{\ast }}{\partial \alpha }>0$ is $%
\alpha <\frac{1-\beta }{2}$ since $\frac{\partial \overline{m}^{\ast }}{%
\partial \alpha }>0\Leftrightarrow 2>3\alpha +\beta ;\frac{\partial 
\overline{r}^{\ast }}{\partial \alpha }>0\Leftrightarrow 1>2\alpha +\beta .$
Again, these more nuanced results are related to the fact that $\alpha $
increases $k^{\ast }$directly, incentivizing investments in $r^{\ast }$and
through it increases in $\overline{m}^{\ast }$. But when $\alpha $ is too
high, the hit to profits can reverse those investments, which hurts the
utility.

We now turn to the firm's profits given by (\ref{firm objective}). In
Appendix \ref{profit final} we show that steady states profits after
introducing $m^{\ast },r^{\ast },$ $k^{\ast }$ and $e^{\ast }$ can be
written as:%
\begin{eqnarray}
\Pi ^{\ast } &=&\frac{\left( 1-\alpha \right) }{1-\delta }\frac{\alpha }{%
A_{e}}\left( \int b_{j}^{2}\right) +\frac{\left( 1-\alpha \right) ^{2}\left(
\alpha +\beta \right) ^{2}}{\left( 1-\delta \right) 4CA_{k}^{2}}
\label{steady state profits} \\
&&\times \left( \frac{1}{2}\left( \int b_{j}^{\frac{1}{3}}\right) ^{4}\left(
\int b_{j}^{\frac{4}{3}}\right) ^{2}+\frac{\lambda \left( 1+2\delta -3\delta
\lambda \right) }{\left( 1-\lambda \right) \left( 1-\delta \lambda \right) }%
\left( \int b_{j}^{\frac{1}{3}}\right) ^{5}\int b_{j}\int b_{j}^{\frac{4}{3}%
}\right.  \notag \\
&&\left. +\frac{\lambda ^{2}\delta \left( 2-\delta -\lambda \delta \right) }{%
2\left( 1-\delta \lambda \right) ^{2}\left( 1-\lambda \right) }\left( \int
b_{j}^{\frac{1}{3}}\right) ^{6}\left( \int b_{j}\right) ^{2}\right)  \notag
\end{eqnarray}

This allows us to perform comparative statics on the parameters.

\begin{proposition}
\label{comparative static pi*}$\Pi ^{\ast }$ is increasing in $\lambda
,\beta ,\delta$. A sufficient condition for $\Pi ^{\ast }$ to be increasing
in $\alpha $ is that $\left( 1-2\alpha -\beta \right) >0,$ and a sufficient
condition for $\Pi ^{\ast }$ to be decreasing in $\alpha $ is that $%
1-2\alpha <0. \Pi ^{\ast }$ is decreasing in $A_{e}$ and $A_{k}$. It is also
decreasing in $C.$
\end{proposition}

\begin{proof}
See the Appendix.
\end{proof}

The effects of the different variables are, as for $u^{\ast },$ inherited
from those of $\overline{m}^{\ast }$, $r^{\ast }$ and $k^{\ast }$ and their
signs have very similar intuitions, adapted to the different functional form
of profits.

\section{Comparative statics with respect to the talent distribution}

To derive comparative statics with respect to the talent distribution, we
will consider what happens when workers become more skilled as represented
by a first-order stochastic dominance shift or when dispersion changes which
can be captured by second-order stochastic dominance. The following Lemma
will be crucial for establishing the results.

\begin{lemma}
\label{SD}a) Let $X$ and $Y$ be real-valued random variables with cumulative
distribution functions $F_{X}$ and $F_{Y}$. Assume that $Y$ \emph{%
first-order stochastically dominates} $X$. for every exponent $p\in
(0,\infty )$ 
\begin{equation*}
\mathbb{E}\!\left( Y^{\,p}\right) \;\geq \;\mathbb{E}\!\left( X^{\,p}\right)
.
\end{equation*}

b) Let $X$ and $Y$ be real-valued random variables with cumulative
distribution functions $F_{X}$ and $F_{Y}$. Assume that $Y$ \emph{%
second-order stochastically dominates} $X$, i.e. 
\begin{equation*}
\int_{-\infty }^{t}F_{Y}(s)\,ds\;\leq \;\int_{-\infty }^{t}F_{X}(s)\,ds\quad 
\text{for every }t\in \mathbb{R}.
\end{equation*}%
Suppose further that $X$ and $Y$ are integrable and non-negative a.s. Then
for every exponent $p\in (0,1)$ 
\begin{equation*}
\mathbb{E}\!\left( Y^{\,p}\right) \;\geq \;\mathbb{E}\!\left( X^{\,p}\right)
.
\end{equation*}%
and for then for every exponent $p\in (1,\infty )$%
\begin{equation*}
\mathbb{E}\!\left( Y^{\,p}\right) \;\leq \;\mathbb{E}\!\left( X^{\,p}\right)
.
\end{equation*}
\end{lemma}

\begin{proof}
a) First-order stochastic dominances (FOSD) is equivalent to 
\begin{equation*}
\mathbb{E}\left( u(Y)\right) \;\geq \;\mathbb{E}\left( u(X)\right) \quad 
\text{for every increasing }u.
\end{equation*}%
For $0<p<\infty $ the map $u_{p}(x)=x^{p}$ is increasing on $\mathbb{R}_{+}$%
, hence the results follows immediately.

b) Second -order dominance (SOSD) is equivalent to the inequality 
\begin{equation*}
\mathbb{E}\left( u(Y)\right) \;\geq \;\mathbb{E}\left( u(X)\right) \quad 
\text{for every increasing, concave }u.
\end{equation*}%
For $0<p<1$ the map $u_{p}(x)=x^{p}$ is both increasing and concave on $%
\mathbb{R}_{+}$, hence the first inequality follows immediately. We also
have that 
\begin{equation*}
\mathbb{E}\left( u(Y)\right) \;\leq \;\mathbb{E}u(X)\quad \text{for every
increasing, convex }u.
\end{equation*}%
For $1<p<\infty $ the map $u_{p}(x)=x^{p}$ is both increasing and convex on $%
\mathbb{R}_{+}$, hence the second inequality follows immediately.
\end{proof}

Since the expressions for $\overline{m}^{\ast }$, $r^{\ast }$, $k^{\ast }$, $%
u^{\ast }$ and $\Pi ^{\ast }$are all increasing functions of $\int b_{j}^{p}$
functions, the following Corollary is immediate from Lemma \ref{SD}.

\begin{corollary}
If the distribution of $b_{j}$ experiences a FOSD shift, $\overline{m}^{\ast
}$, $r^{\ast }$, $k^{\ast }$, $u^{\ast }$ and $\Pi ^{\ast }$increase.
\end{corollary}

In other words, if the talent distribution shifts upwards (i.e. workers
become more skilled) meaning, purpose, socialization, utility and profits
all increase.

\begin{empirical implication}
Investing in talent development (e.g., training) enhances the returns to
purpose-driven strategies.
\end{empirical implication}

Empirical evidence supports the idea that investment in purpose is closely
linked to firms' broader human capital strategies. For instance, the SHRM \&
TalentLMS (2022) Workplace Learning and Development Trends Report\footnote{%
https://www.shrm.org/content/dam/en/shrm/research/2022-Workplace-Learning-and-Development-Trends-Report.pdf%
} finds that 83\% of HR managers believe training helps attract talent, and
76\% of employees are more likely to stay with a company that offers
continuous learning---suggesting that educational investment is perceived as
part of a meaningful work environment. Similarly, empirical evidence from
Harvard Business Publishing (2025) confirms that purpose-driven
organizations systematically integrate educational investment into their
talent strategies. McKinsey (2024) further emphasizes that purpose-driven
transformation programs often include capability building and leadership
development as core components, reinforcing the strategic link between
purpose and talent.

The situation is more nuanced when the talent distributions experience a
SOSD shift. The $\int b_{j}^{p}$ functions have some exponents $p<1$ and
others $p>1.$ However, we can get a definite sign for the change when we
make assumptions on what the baseline distribution is.

\begin{proposition}
\label{SOSD Prop}If the distribution of $b_{j}^{\frac{1}{3}}$ undergoes a
SOSD shift, then, $\overline{m}^{\ast }$, $r^{\ast }$, $k^{\ast }$, $u^{\ast
}$ and $\Pi ^{\ast }$decrease. If the distribution of $b_{j}^{2}$ undergoes
a SOSD shift, then, $\overline{m}^{\ast }$, $r^{\ast }$, $k^{\ast }$, $%
u^{\ast }$ and $\Pi ^{\ast }$increase.

\begin{proof}
The result follow from Lemma \ref{SD}, by noting that all the $\int
b_{j}^{p} $ expressions in $\overline{m}^{\ast }$, $r^{\ast }$, $k^{\ast }$, 
$u^{\ast } $ and $\Pi ^{\ast }$ integrate convex transformations of $b_{j}^{%
\frac{1}{3}} $ and concave transformations of $b_{j}^{2}.$
\end{proof}
\end{proposition}

In Appendix \ref{examples lognormal} we discuss a couple of relevant cases
including an intermediate one, using the lognormal distribution, to get more
intuition.

To understand what is going on in this problem, remember that individual
output can be written as:

\begin{equation}
z_{i,t}\left( m_{i,t},e_{i,t}\right) =b_{i,t}\left( b_{i,t}\frac{\alpha }{%
A_{e}}+\frac{\alpha +\beta }{2A_{k}}r_{t}\left( b_{i,t}^{\frac{1}{3}}\right)
\left( \int b_{j,t}^{\frac{1}{3}}\right) ^{2}+\lambda \overline{m}%
_{t-1}\right)  \label{individual output 1}
\end{equation}

Notice that in the equation (\ref{individual output 1}) for individual
output the individual talent parameter $b_{i,t}$ is squared from the first
term and to the power $4/3$ from the second term. When this is aggregated to
the total output of the firm this means we will have terms with $\int
b_{j,t}^{\frac{4}{3}}$ and $\int b_{j,t}^{2},$ in addition to $\int b_{j,t}^{%
\frac{1}{3}}$ which comes from $k_{t}$ and the $\overline{m}_{t-1}$ also has
a $\int b_{j,t}$ term. These different terms have different reactions to the
decrease in dispersion as per Lemma \ref{SD}, which is why the Proposition %
\ref{SOSD Prop} depends on whether the distributions of the random variables 
$b_{j,t}^{2},$ or $b_{j,t}^{\frac{1}{3}}$ are the baseline.

Notice that when the support of the distribution starts at 1, if the
baseline distribution is for $b_{j}^{\frac{1}{3}},$ it is a more
concentrated distribution than for $b_{j}^{p}$ for a higher power. So a SOSD
shift decreases dispersion from a lower base. On the contrary, if the
baseline distribution is for $b_{j}^{2}$ it is a less concentrated
distribution than for $b_{j}^{p}$ for a lower power. So a SOSD shift
decreases dispersion from a higher base. This suggests that increasing
dispersion will increase $\overline{m}^{\ast }$, $r^{\ast }$, $k^{\ast }$, $%
u^{\ast }$ and $\Pi ^{\ast }$ when the starting point is not too disperse,
and will decrease $\overline{m}^{\ast }$, $r^{\ast }$, $k^{\ast }$, $u^{\ast
}$ and $\Pi ^{\ast }$ when the starting distribution is more disperse. We
can conclude that

\begin{empirical implication}
\label{intermediate dispersion}Some intermediate level of heterogeneity in
skills seems to be best for performance.
\end{empirical implication}

Companies around the world are now more sensitive to diversity in talents.
Usually this means that they try to foster an environment in which different
ideas and perspectives are valued, perhaps because they can contribute to
generate new lines of business, or more productive project. The kind of
diversity we contemplate here is different in nature. Everything else equal,
some workers are more productive than others. In a sense, we are talking
about \emph{vertical} differentiation, whereas the more standard discourse
is about \emph{horizontal }diversity. Because of the way output (as shown in
equation (\ref{individual output 1})) interacts with the distribution of
talent in potentially complicated ways, the \emph{vertical} differentiation
also matters, and companies need to pay attention to it.

To our knowledge, no study directly tests the empirical implication \ref%
{intermediate dispersion} that intermediate levels of skill heterogeneity
maximize firm performance by enhancing employees' sense of meaning and
purpose. The closest evidence is provided by Iranzo et al. (2008), who
analyze employer--employee matched data from Italian manufacturing firms.
They construct a latent skill index from wage regressions and quantify
within-firm skill dispersion. Their results indicate that productivity
increases with dispersion within occupational groups (e.g., among production
workers) but decreases when dispersion occurs across groups (e.g., between
production and non-production workers). This pattern implies a nonmonotonic
relationship between skill heterogeneity and performance---structured
diversity within roles is beneficial, whereas excessive dispersion across
roles is detrimental. Nevertheless, these findings do not constitute a
direct test of our proposed mechanism, and further research is needed.

\section{Path to steady state}

In our reformulation of the firm's problem, the firm maximizes the objective
function $\max_{\left\{ \overline{m}_{t}\right\} _{t=0}^{\infty
}}\sum_{t=0}^{\infty }\delta ^{t-1}F(\overline{m}_{t-1},\overline{m}%
_{_{t}},z)$. An inspection of $F(\overline{m}_{t-1},\overline{m}_{_{t}},z) $
shows that this function is quadratic in $\overline{m}_{t-1}$ and $\overline{%
m}_{t}$. Since the problem is quadratic we know from Anton et al. (1998)
that the roots of the characteristic equation determine the approach to the
steady state. In Appendix \ref{roots}\ we show that 
\begin{equation*}
\overline{m}_{t}=\overline{m}^{\ast }+\mu _{2}^{t}\left( \overline{m}_{0}-%
\overline{m}^{\ast }\right)
\end{equation*}%
with 
\begin{equation*}
\mu _{2}=\frac{1}{2}\left( \frac{1+\delta \lambda ^{2}}{\delta \lambda }-%
\sqrt{\left( \frac{1+\delta \lambda ^{2}}{\delta \lambda }\right) ^{2}-\frac{%
4}{\delta }}\right)
\end{equation*}

Note that $\mu _{2}<1$ iff $\delta \lambda <1$, so the solution approaches
the steady state monotonically from the initial condition.

\section{The worker-owned firm}

While both workers and firms benefit from operating in an environment imbued
with positive meaning or purpose, this alignment does not necessarily imply
a convergence of interests or the absence of tensions regarding the value
each party places on such meaning. One way to illustrate this divergence is
by comparing the decisions made by a for-profit firm with those that might
be expected from a firm that is entirely owned and operated by its workers.
In the worker-owned firm, clearly, the workers would keep all the output, so 
$\alpha =1$ in the worker utility (\ref{Utility worker}), but also, since
the workers are now the firm $1-\alpha =1$ in the objective function of the
firm (\ref{firm objective}). To be clear, $\alpha $ is renormalized
independently in (\ref{Utility worker}) and in (\ref{firm objective}). Thus
the solutions to the worker-owned firm in steady state are as follows:

\begin{itemize}
\item for meaning $\overline{m}_{wo}^{\ast }$%
\begin{equation*}
\overline{m}_{wo}^{\ast }=\frac{\int b_{j}^{\frac{4}{3}}\left( 1+\beta
\right) ^{2}\left( \int b_{j}^{\frac{1}{3}}\right) ^{5}}{C4A_{k}^{2}\left(
1-\lambda \right) }+\frac{\delta \lambda \int b_{j}\left( 1+\beta \right)
^{2}\left( \int b_{j}^{\frac{1}{3}}\right) ^{6}}{C\left( 1-\delta \lambda
\right) 4A_{k}^{2}\left( 1-\lambda \right) }
\end{equation*}

\item for steady state investment in purpose $r_{wo}^{\ast }$ 
\begin{equation*}
r_{wo}^{\ast }=\frac{\int b_{j,t}^{\frac{4}{3}}\left( 1+\beta \right) \left(
\int b_{j,t}^{\frac{1}{3}}\right) ^{2}}{2CA_{k}}+\frac{\delta \lambda \int
b_{j,t}\left( 1+\beta \right) \left( \int b_{j,t}^{\frac{1}{3}}\right) ^{3}}{%
2CA_{k}\left( 1-\delta \lambda \right) }
\end{equation*}

\item for socialization $k_{wo}^{\ast }$
\end{itemize}

\begin{equation*}
k_{wo}^{\ast }=\frac{1+\beta }{2A_{k}}\left( \int b_{j}^{\frac{1}{3}}\right)
\left( \int b_{j}^{\frac{2}{3}}\right) \left( \frac{\int b_{j}^{\frac{4}{3}%
}\left( 1+\beta \right) \left( \int b_{j}^{\frac{1}{3}}\right) ^{2}}{2CA_{k}}%
+\frac{\delta \lambda \int b_{j}\left( 1+\beta \right) \left( \int b_{j}^{%
\frac{1}{3}}\right) ^{3}}{2CA_{k}\left( 1-\delta \lambda \right) }\right) ^{%
\frac{1}{2}}
\end{equation*}%
Simple inspection shows that $k_{wo}^{\ast }\geq k^{\ast }$, $r_{wo}^{\ast
}\geq r^{\ast },$ and $\overline{m}_{wo}^{\ast }\geq \overline{m}^{\ast }$
which implies that there will be some kind of friction between management
and workers about this issue. The firm invests less in purpose than the
worker-owned firm would, resulting in a lower common component for
socialization and in lower meaning.

\begin{empirical implication}
Worker owned firms invest more in purpose and meaning than traditional firms.
\end{empirical implication}

Evidence suggests that worker-owned firms tend to institutionalize practices
that reinforce organizational purpose and employee meaning more than
conventional firms, though most studies are descriptive. Case studies of
Mondrag\'{o}n, a group of 96 cooperatives in the Basque Country, show that
worker ownership is tied to strong investments in socialization and
meaning-building---through cooperative education programs, participatory
governance structures, and initiatives to transmit cooperative values across
contexts (Flecha and Ngai, 2014; Morl\`{a}-Folch et al., 2021).
Meta-analyses show that employee ownership is linked to higher
organizational commitment (Wagner et al., 2015) and modest improvements in
firm performance (O'Boyle et al., 2016).

\section{Conclusions}

We have developed a dynamic model in which a profit-maximizing firm invests
in workplace purpose to enhance the motivation and effort of workers who
value meaning in their jobs. In our framework, meaning emerges endogenously
from the interaction between firm investment and workers' socialization
effort, and it directly contributes to both worker utility and firm
productivity. A central result is that meaning and purpose are increasing in
the workers' share of output, but only up to a point: when the share is
moderate, meaning and profit incentives are complementary; when the share
becomes too high, they act as substitutes. This highlights a subtle
trade-off in the use of monetary and non-monetary incentives to motivate
workers. Another key finding is that some intermediate level of
heterogeneity in skills seems to be best for performance leading to higher
levels of purpose, socialization, productivity, individual utility and
firm's profits. This provides a theoretical rationale for the increasing
emphasis on choosing the correct mix of diversity in firms and
organizations, particularly those aiming to foster purpose-driven cultures.

Several directions for future research naturally emerge from our analysis.
First, we have treated workers' concern for meaning as exogenous, whereas in
practice firms may influence this through hiring policies or by cultivating
a preference for meaning internally. Selective recruitment strategies or
internal messaging efforts could be incorporated into the model to study how
firms shape preferences over time. Second, while our model allows for
heterogeneity in worker ability, all workers are assumed to care about
meaning to the same degree. Introducing heterogeneity in preferences for
meaning, alongside differences in productivity, would be relevant in
environments where firms differ in their specialization and may wish to
attract specific types of employees. Although our model focuses on a single
representative firm, extending it to a setting with multiple firms and a
competitive labor market would allow for a richer analysis of sorting and
matching. Finally, future work could consider multi-dimensional
heterogeneity, where workers differ not only in their overall productivity
but also in how effective they are at productive tasks versus socialization.
This would provide a deeper understanding of how organizations allocate
roles and design incentives in the presence of complex trade-offs between
purpose and performance.\newpage

\section*{References}
 Albornoz, F., Cabrales, A., \& Hauk, E. (2019). Occupational
choice with endogenous spillovers. The Economic Journal, 129(621), 1953-1970.

 Allan, B. A., Batz-Barbarich, C., Sterling, H. M., \& Tay, L.
(2019). Outcomes of meaningful work: A meta-analysis. Journal of management studies, 56(3), 500-528.

 Allan, B. A., Dexter, C., Kinsey, R., \& Parker, S. (2018).
Meaningful work and mental health: job satisfaction as a moderator. Journal
of mental health, 27(1), 38-44.

 Ant\'{o}n, J., Cerd\'{a}, E. \& Huergo (1998), E. Sensitivity
analysis in A class of dynamic optimization models. Top 6, 97--121 .
https://doi.org/10.1007/BF02564800

 Anwar, A., Waqas, A., Shakeel, K., \& Hassan, S. (2018). Impact
of intrinsic and extrinsic motivation on employee's retention: A case from
call center. International Journal of Academic Research in Business and
Social Sciences, 8(6), 652--666.

 Armouti-Hansen, J., Cassar, L., Dereky, A., \& Engl, F. (2024).
Efficiency wages with motivated agents. Games and Economic Behavior, 145,
66-83.

 Ashraf, N., Bandiera, O., Minni, V., \& Zingales, L. (2025).
Meaning at work (No. w33843). National Bureau of Economic Research.

 Besley, T. \& Ghatak, M. (2005). Competition and incentives with
motivated agents', American Economic Review, vol. 95(3), pp. 616--36.

 Blount, S., \& Leinwand, P. (2019). Why are we here?: If you want
employees who are more engaged and productive, give them a purpose---one
concretely tied to your customers and your strategy. Harvard business
review, 2019(November-December), 1-9.

 Bo\v{s}tjan\v{c}i\v{c}, E., Antolovi\'{c}, S., \& Er\v{c}ulj, V.
(2018). Corporate volunteering: Relationship to job resources and work
engagement. Frontiers in psychology, 9, 1884.

 Burbano, V. C. (2016). Social responsibility messages and worker
wage requirements: Field experimental evidence from online labor
marketplaces. Organization Science, 27(4), 1010-1028.

 Cabrales, A., Calv\'{o}-Armengol, A., \& Zenou, Y. (2011). Social
interactions and spillovers. Games and Economic Behavior, 72(2), 339-360.

 Cassar, L. \& Armouti-Hansen, J., (2020). Optimal contracting
with endogenous project mission. J. Eur. Econ. Assoc. 18 (5), 2647--2676.

 Cassar, L., \& Meier, S. (2018). Nonmonetary incentives and the
implications of work as a source of meaning. Journal of Economic
Perspectives, 32(3), 215-238.

 Cnossen, F., \& Nikolova, M. (2024). Work Meaningfulness and
Effort (No. 17182). IZA Discussion Papers

 Delfgaauw, J., \& Dur, R. (2007). Signaling and screening of
workers' motivation. Journal of Economic Behavior \& Organization, 62(4),
605-624.

 Delfgaauw, J., \& Dur, R. (2008). Incentives and workers'
motivation in the public sector. The Economic Journal, 118(525), 171-191.

 Dik, B. J., Duffy, R. D., \& Steger, M. F. (2012). Enhancing
social justice by promoting prosocial values in career development
interventions. Counseling and Values, 57, 31--37.

 Fang, Q., Wen, C., \& Xu, H. (2024). Long-term oriented culture,
performance pressure and corporate innovation: Evidence from China. Plos
one, 19(5), e0302148.

 Farrell, N. (2024, November 21). Leveraging the seven intrinsic
motivators for call center success in 2025. AnswerNet. Retrieved 4 December
2025.
https://answernet.com/leveraging-the-seven-intrinsic-motivators-for-call-center-success-in-2025/

 Flammer, C., \& Bansal, P. (2017). Does a long-term orientation
create value? Evidence from a regression discontinuity. Strategic Management
Journal, 38(9), 1827-1847

 Flecha, R., \& Ngai, P. (2014). The challenge for Mondragon:
Searching for the cooperative values in times of internationalization.
Organization, 21(5), 666-682.

 Gartenberg, C. (2023) The Contingent Relationship Between Purpose
and Profits. Strategy Science 8(2):256-269.
https://doi.org/10.1287/stsc.2023.0194

 Gartenberg, C., Prat, A., \& Serafeim, G. (2019). Corporate
purpose and financial performance. Organization Science, 30(1), 1-18.

 George, G., Haas, M. R., McGahan, A. M., Schillebeeckx, S. J., \&
Tracey, P. (2023). Purpose in the for-profit firm: A review and framework
for management research. Journal of management, 49(6), 1841-1869.

 Gitnux. (2025). HR in the retail industry statistics: Market data
report 2025. Retrieved October 20, 2025.

https://gitnux.org/hr-in-the-retail-industry-statistics

 Grueso Hinestroza, M. P., Ant\'{o}n, C., \& L\'{o}pez-Santamar%
\'{\i}a, M. (2022). Meaningful work and satisfaction with life: A case study
from a supported employment program---Colombia. Behavioral Sciences, 12(7),
229.

 Harvard Business Publishing. (2025). The Purpose Factor: Why Your
Talent Strategy (and So Much More) Depends on It. Retrieved October 20, 2025

https://www.harvardbusiness.org/wp-content/uploads/2025/05/CRE6406\_CL\_PER%
\_The-Purpose-Factor.pdf

 Insignia Resources. (2025). Call center turnover rates: 2025
industry average. Retrieved October 30, 2025.
https://www.insigniaresource.com/research/call-center-turnover-rates/

 Iranzo, S., Schivardi, F., \& Tosetti, E. (2008). Skill
dispersion and firm productivity: An analysis with employer-employee matched
data. Journal of Labor Economics, 26(2), 247-285.

 Jasinenko, A., \& Steuber, J. (2023). Perceived organizational
purpose: Systematic literature review, construct definition, measurement and
potential employee outcomes. Journal of Management Studies, 60(6), 1415-1447.

 Kesternich, I., Schumacher, H., Siflinger, B., \& Schwarz, S.
(2021). Money or meaning? Labor supply responses to work meaning of employed
and unemployed individuals. European Economic Review, 137, 103786.

 Knox, B. D. (2020). Employee volunteer programs are associated
with firm-level benefits and CEO incentives: Data on the ethical dilemma of
corporate social responsibility activities. Journal of Business Ethics,
162(2), 449-472.

 McKinsey \& Company. (2024). Insights to shape organization
culture for success. Retrieved October 30, 2025.
https://www.mckinsey.com/capabilities/people-and-organizational-performance/our-insights/the-organization-blog/insights-to-shape-organization-culture-for-success

 Mercer. (2025). Results of the 2025 US turnover surveys.
Retrieved October 30, 2025.
https://www.imercer.com/articleinsights/workforce-turnover-trends

 Morl\`{a}-Folch, T., Aubert Simon, A., de Freitas, A. B., \& Hern%
\'{a}ndez-Lara, A. B. (2021). The Mondragon case: Companies addressing
social impact and dialogic methodologies. International Journal of
Qualitative Methods, 20, 1--9.

 Nikolova, M., \& Cnossen, F. (2020). What makes work meaningful
and why economists should care about it. Labour economics, 65, 101847.

 O'Boyle, E. H., Patel, P. C., \& Gonzalez-Mul\'{e}, E. (2016).
Employee ownership and firm performance: a meta-analysis. Human Resource
Management Journal, 26(4), 425-448.

 Reker, G. T., Peacock, E. J., \& Wong, P. T. P. (1987). Meaning
and purpose in life and well-being: A life-span perspective. Journal of
Gerontology, 42, 44.

 Schnell, T. (2009). The Sources of Meaning and Meaning in Life
Questionnaire (SoMe): Relations to demographics and well-being. Journal of
Positive Psychology, 4, 483--499.

 SHRM \& TalentLMS. (2022). Workplace Learning and Development
Trends Report. Society for Human Resource Management. Retrieved October 30,
2025.
https://www.shrm.org/content/dam/en/shrm/research/2022-Workplace-Learning-and-Development-Trends-Report.pdf

 Steger, M. F. (2016). Creating meaning and purpose at work. The
Wiley Blackwell handbook of the psychology of positivity and strengths-based
approaches at work, 60-81.

 Steger, M. F., Kashdan, T. B., \& Oishi, S. (2008). Being good by
doing good: Eudaemonic activity and daily well-being correlates, mediators,
and temporal relations. Journal of Research in Personality, 42, 22--42.

 Vazquez, C. A. (2025, March 5). Motivating tech teams: Theories
\& practical strategies. Retrieved October 30, 2025
https://coderslink.com/employers/blog/the-psychology-behind-tech-worker-motivation-from-herzberg-to-hybrid/

 Veleva, V., Parker, S., Lee, A., \& Pinney, C. (2012). Measuring
the business impacts of community involvement: The case of employee
volunteering at UL. Business and Society Review, 117(1), 123-142.

 Wagner, S. H., Bailey, E., Bush, C., \& Filipkowski, M. (2015).
Employee ownership and organizational commitment: A meta-analysis. Institute
for the Study of Employee Ownership and Profit Sharing, Rutgers University.

 Wandycz-Mejias, J., Rold\'{a}n, J. L., \& Lopez-Cabrales, A.
(2025). Analyzing the impact of work meaningfulness on turnover intentions
and job satisfaction: A self-determination theory perspective. Journal of
Management \& Organization, 31(1), 384-407.

 Wang, T., Zhao, B., \& Thornhill, S. (2015). Pay dispersion and
organizational innovation: The mediation effects of employee participation
and voluntary turnover. Human relations, 68(7), 1155-1181.

 Work Institute. (2025). 2025 retention report: Employee retention
truths in today's workplace. Retrieved October 30, 2025.
https://workinstitute.com/retention-reports

\appendix

\section*{Appendix}

\section{\textbf{Proof of Lemma \protect\ref{lemma worker problem} }}

We conjecture that 
\begin{equation*}
e_{i,t}=b_{i,t}^{\rho }e_{t};k_{i,t}=b_{i,t}^{\eta }k
\end{equation*}%
and show that it indeed holds. We introduce this conjecture into (\ref{FOCe}%
) and (\ref{FOCk}) and get%
\begin{equation*}
\alpha b_{i,t}-A_{e}b_{i,t}^{\rho }e_{t}=0
\end{equation*}%
and%
\begin{equation*}
\frac{1}{2}\left( \alpha +\beta \right) b_{i,t}\left( r_{t}^{\frac{1}{2}%
}\left( b_{i,t}^{\eta \left( -\frac{1}{2}\right) }k^{-\frac{1}{2}}\int
b_{j,t}^{\frac{\eta }{2}}k^{\frac{1}{2}}\right) \right) -b_{i,t}^{\eta
}A_{k}k=0
\end{equation*}%
respectively. These new conditions coincide with the original first order
conditions if 
\begin{eqnarray*}
1 &=&\rho \\
1+\eta \left( -\frac{1}{2}\right) &=&\eta \\
\eta &=&\frac{2}{3}
\end{eqnarray*}%
Hence $k_{t}=\frac{\alpha +\beta }{2A_{k}}r_{t}^{\frac{1}{2}}\left( \int
b_{j,t}^{\frac{1}{3}}\right) $ and $e_{t}=\frac{\alpha }{A_{e}}$ .

\section{\textbf{Proof of Proposition \protect\ref{steady state values}}}

Isolating $r_{t}$ from the law of motion (\ref{law of motion}) we get

\begin{equation}
r_{t}=\frac{2A_{k}\left( \overline{m}_{t}-\lambda \overline{m}_{t-1}\right) 
}{\left( \alpha +\beta \right) \left( \int b_{j,t}^{\frac{1}{3}}\right) ^{3}}%
\geq 0  \label{r_t}
\end{equation}%
Since $r_{t}\geq 0$ it is also true that $\overline{m}_{t}\geq \lambda 
\overline{m}_{t-1}$.

We now reformulate our problem using Anton et al. (1998) and write the
maximization problem as a function of \ $\overline{m}_{t-1},\overline{m}%
_{_{t}}$. Our problem therefore becomes 
\begin{eqnarray*}
&&\max_{\left\{ \overline{m}_{t}\right\} _{t=0}^{\infty }}\sum_{t=0}^{\infty
}\delta ^{t-1}F(\overline{m}_{t-1},\overline{m}_{_{t}},z) \\
&&\text{such that }\overline{m}_{t}\geq \lambda \overline{m}_{t-1} \\
&&\text{and }\overline{m}_{o}\text{ given}
\end{eqnarray*}%
where $z$ refers to any of our exogenous parameters in the model and $F(%
\overline{m}_{t-1},\overline{m}_{t},z)$ is a quadratic function in $%
\overline{m}_{t}$ and $\overline{m}_{t-1}$ given by 
\begin{eqnarray}
F(\overline{m}_{t-1},\overline{m}_{t},z) &=&\left( 1-\alpha \right) \left(
\int b_{j,t}^{2}\frac{\alpha }{A_{e}}+\frac{\left( \overline{m}%
_{_{t}}-\lambda \overline{m}_{t-1}\right) }{\int b_{j,t}^{\frac{1}{3}}}%
\left( \int b_{j,t}^{\frac{4}{3}}\right) +\lambda \int b_{j,t}\overline{m}%
_{t-1}\right)  \notag \\
&&-\frac{C}{2}\left( \frac{2A_{k}\left( \overline{m}_{t}-\lambda \overline{m}%
_{t-1}\right) }{\left( \alpha +\beta \right) \left( \int b_{j,t}^{\frac{1}{3}%
}\right) ^{3}}\right) ^{2}  \label{Capital F function}
\end{eqnarray}

From now on we assume the type distribution is stationary, and thus all the $%
\int \left( b_{j,t+1}\right) ^{x}=\int \left( b_{j}\right) ^{x}.$

In order to apply the solution techniques developed in Anton et al. (1998)
we need to make sure that the required assumptions stated in the paper hold.
We show in section \ref{Anton assumptions} that these assumptions indeed
hold if we assume that 
\begin{equation}
-\frac{\int b_{j}^{\frac{4}{3}}}{\int b_{j}^{\frac{1}{3}}}+\int b_{j}>0
\label{A3}
\end{equation}%
which we do from now on.

The Euler equation of this maximization is 
\begin{equation}
0=F_{\overline{m}_{t}}(\overline{m}_{t-1},\overline{m}_{t,}z)+\delta F_{%
\overline{m}_{t}}(\overline{m}_{t},\overline{m}_{t+1,}z)  \label{Euler}
\end{equation}%
And the transversality condition is given by 
\begin{equation}
\lim_{t\rightarrow \infty }\delta ^{t}F_{\overline{m}_{t-1}}(\overline{m}%
_{t-1},\overline{m}_{t,}z)\overline{m}_{t-1}=0  \label{Transversality}
\end{equation}

We need to use the Euler equation at steady state where $\overline{m}^{\ast
}=\overline{m}_{t-1}=\overline{m}_{t}=\overline{m}_{t+1}$. While $F(%
\overline{m}_{t-1},\overline{m}_{t,}z)$ is given by equation (\ref{Capital F
function}), we also need to define $F(\overline{m}_{t},\overline{m}_{t+1,}z)$
as

\begin{eqnarray*}
F(\overline{m}_{t},\overline{m}_{t+1,}z) &=&\left( 1-\alpha \right) \left(
\int b_{j,t+1}^{2}\frac{\alpha }{A_{e}}+\frac{\left( \overline{m}%
_{_{t}+1}-\lambda \overline{m}_{t}\right) }{\int b_{j,t+1}^{\frac{1}{3}}}%
\left( \int b_{j,t+1}^{\frac{4}{3}}\right) +\lambda \int b_{j,t}\overline{m}%
_{t}\right) \\
&&-\frac{C}{2}\left( \frac{2A_{k}\left( \overline{m}_{t+1}-\lambda \overline{%
m}_{t}\right) }{\left( \alpha +\beta \right) \left( \int b_{j,t+1}^{\frac{1}{%
3}}\right) ^{3}}\right) ^{2}
\end{eqnarray*}%
Therefore 
\begin{eqnarray*}
F_{\overline{m}_{t}}(\overline{m}_{t},\overline{m}_{t+1,}z) &=&\frac{%
\partial F(\overline{m}_{t},\overline{m}_{t+1,}z)}{\partial \overline{m}_{t,}%
}=\left( 1-\alpha \right) \left( -\frac{\lambda \int b_{j,t+1}^{\frac{4}{3}}%
}{\int b_{j,t+1}^{\frac{1}{3}}}+\lambda \int b_{j,t+1}\right) \\
&&+C\frac{4A_{k}^{2}\lambda \left( \overline{m}_{t+1}-\lambda \overline{m}%
_{t}\right) }{\left( \alpha +\beta \right) ^{2}\left( \int b_{j,t+1}^{\frac{1%
}{3}}\right) ^{6}}
\end{eqnarray*}%
and 
\begin{equation*}
F_{\overline{m}_{t}}(\overline{m}_{t-1},\overline{m}_{t,}z)=\frac{\partial F(%
\overline{m}_{t-1},\overline{m}_{t,}z)}{\partial \overline{m}_{t,}}=\left(
1-\alpha \right) \left( \frac{\int b_{j,t}^{\frac{4}{3}}}{\int b_{j,t}^{%
\frac{1}{3}}}\right) -C\frac{4A_{k}^{2}\left( \overline{m}_{t}-\lambda 
\overline{m}_{t-1}\right) }{\left( \alpha +\beta \right) ^{2}\left( \int
b_{j,t}^{\frac{1}{3}}\right) ^{6}}
\end{equation*}%
We assume the type distribution is stationary, and thus all the $\int \left(
b_{j,t+1}\right) ^{x}=\int \left( b_{j}\right) ^{x}.$

Using a stationary type distribution in the Euler equation at steady state
where $\overline{m}^{\ast }=\overline{m}_{t-1}=\overline{m}_{t}=\overline{m}%
_{t+1,}$ we get%
{\footnotesize \begin{equation*}
\left( 1-\alpha \right) \left( \frac{\int b_{j}^{\frac{4}{3}}}{\int b_{j}^{%
\frac{1}{3}}}\right) -C\frac{4A_{k}^{2}\left( \left( 1-\lambda \right) 
\overline{m}^{\ast }\right) }{\left( \alpha +\beta \right) ^{2}\left( \int
b_{j}^{\frac{1}{3}}\right) ^{6}}+\delta \left( \left( 1-\alpha \right)
\left( -\frac{\lambda \int b_{j}^{\frac{4}{3}}}{\int b_{j}^{\frac{1}{3}}}%
+\lambda \int b_{j}\right) +C\frac{4A_{k}^{2}\lambda \left( 1-\lambda
\right) \overline{m}^{\ast }}{\left( \alpha +\beta \right) ^{2}\left( \int
b_{j}^{\frac{1}{3}}\right) ^{6}}\right) =0
\end{equation*}%
}
\begin{equation*}
\left( 1-\alpha \right) \left( 1-\delta \lambda \right) \left( \frac{\int
b_{j}^{\frac{4}{3}}}{\int b_{j}^{\frac{1}{3}}}\right) +\left( 1-\alpha
\right) \delta \lambda \int b_{j}=C\left( 1-\delta \lambda \right) \frac{%
4A_{k}^{2}\left( 1-\lambda \right) \overline{m}^{\ast }}{\left( \alpha
+\beta \right) ^{2}\left( \int b_{j}^{\frac{1}{3}}\right) ^{6}}
\end{equation*}%
Isolating $\overline{m}^{\ast }$ we find 
\begin{equation*}
\overline{m}^{\ast }=\frac{\left( 1-\alpha \right) \left( 1-\delta \lambda
\right) \left( \frac{\int b_{j}^{\frac{4}{3}}}{\int b_{j}^{\frac{1}{3}}}%
\right) +\left( 1-\alpha \right) \delta \lambda \int b_{j}}{C\left( 1-\delta
\lambda \right) \frac{4A_{k}^{2}\left( 1-\lambda \right) }{\left( \alpha
+\beta \right) ^{2}\left( \int b_{j}^{\frac{1}{3}}\right) ^{6}}}
\end{equation*}%
which can be rewritten as (\ref{steady state m}). Using (\ref{steady state m}%
) the steady state investment in purpose $r^{\ast }=\frac{2A_{k}\left(
1-\lambda \right) \overline{m}^{\ast }}{\left( \alpha +\beta \right) \left(
\int b_{j,t}^{\frac{1}{3}}\right) ^{3}}$ is expressed in primitives of the
model as (\ref{steady state r}). Introducing (\ref{steady state r}) into $%
k^{\ast }=\frac{\alpha +\beta }{2A_{k}}\left( r^{\ast }\right) ^{\frac{1}{2}%
}\left( \int b_{j}^{\frac{1}{3}}\right) \,$\ as derived in (\ref%
{socialization effort}) delivers (\ref{steady state k}).

We now show that the transversality condition ( \ref{Transversality}) holds.
We need to calculate 
\begin{equation*}
F_{\overline{m}_{t-1}}(\overline{m}_{t-1},\overline{m}_{t,}z)=\left(
1-\alpha \right) \left( -\frac{\lambda \int b_{j}^{\frac{4}{3}}}{\int b_{j}^{%
\frac{1}{3}}}+\lambda \int b_{j}\right) +C\frac{4A_{k}^{2}\lambda \left( 
\overline{m}_{t}-\lambda \overline{m}_{t-1}\right) }{\left( \alpha +\beta
\right) ^{2}\left( \int b_{j}^{\frac{1}{3}}\right) ^{6}}
\end{equation*}%
Since the first term is a constant the transversality condition (\ref%
{Transversality}) will be satisfied if%
\begin{equation*}
\lim_{t\rightarrow \infty }\delta ^{t}\left( \overline{m}_{t}-\lambda 
\overline{m}_{t-1}\right) \overline{m}_{t-1}=0.
\end{equation*}%
This is trivially satisfied in our case, since we focus on a stationary
solution, and $\lim_{t\rightarrow \infty }\overline{m}_{t}=\overline{m}%
^{\ast }$ is a constant.

In section \ref{saddle point}\ we show that the steady state $\overline{m}%
^{\ast }$ exhibits saddle point stability since it satisfies condition (4.7)
in Anton et al. (1998) given by 
\begin{equation}
\frac{\partial ^{2}F}{\partial \overline{m}_{t}^{2}}+\delta \frac{\partial
^{2}F}{\partial \overline{m}_{t-1}^{2}}+\left\vert \frac{\partial ^{2}F}{%
\partial \overline{m}_{t}\partial \overline{m}_{t-1}}\right\vert \left(
1-\delta \right) <0\text{.}  \label{stability}
\end{equation}

\section{\textbf{Checking the conditions in Anton et al. (1998) \label{Anton
assumptions}}}

To apply Anton et al. (1998) the following conditions have to be satisfied.

\begin{description}
\item[A1] $F$ is continuous and continuously differentiable in the interior
of the domain with respect to $(\overline{m}_{t-1},\overline{m}_{t})$, for
each $z$

\item[A2] The domain is a subset of the positive reals.

\item[A3] For each $\overline{m}_{t},z$ a, $F(.,\overline{m}_{t},z)$ is
strictly increasing in its first argument

\item[A4] For each $z$, $F(\overline{m}_{t-1},\overline{m}_{t,}z)$ is
concave in $\overline{m}_{t-1},\overline{m}_{t}$

\item[A5] For each $z$, and $\delta \in (0,1)$, there exists a unique $%
\overline{m}^{\ast }$ satisfying $F_{\overline{m}_{t}}(\overline{m}^{\ast },%
\overline{m}^{\ast },z)+\delta F_{\overline{m}_{t}}(\overline{m}^{\ast },%
\overline{m}^{\ast },z)=0$

\item[A6 ] $F$ is twice continuously differentiable in the domain
\end{description}

A1, A2 and A6 clearly hold. For A3 first we need to check

\begin{eqnarray*}
F_{\overline{m}_{t-1}}(\overline{m}_{t-1},\overline{m}_{t,}z) &=&\frac{%
\partial F(\overline{m}_{t-1},\overline{m}_{t,}z)}{\partial \overline{m}%
_{t-1}} \\
&=&\left( 1-\alpha \right) \lambda \left( -\frac{\int b_{j}^{\frac{4}{3}}}{%
\int b_{j}^{\frac{1}{3}}}+\int b_{j}\right) +C\lambda \frac{4A_{k}^{2}\left( 
\overline{m}_{t}-\lambda \overline{m}_{t-1}\right) }{\left( \alpha +\beta
\right) ^{2}\left( \int b_{j,t+1}^{\frac{1}{3}}\right) ^{6}}
\end{eqnarray*}

We can guarantee that this is positive if we assume condition \ref{A3} given 
$\left( \overline{m}_{t}-\lambda \overline{m}_{t-1}\right) =\overline{s}%
_{_{t}}r_{t}^{\frac{1}{2}}\geq 0,$ so that A3 holds.

A4 is true because the first term $F(\overline{m}_{t-1},\overline{m}_{t,}z)$
is linear in $\overline{m}_{t-1},\overline{m}_{t,}$ and the second term is
strictly concave in $\overline{m}_{t-1},\overline{m}_{t}$.

Our stationary solution to the maximization problem satisfies A5.

\subsection{\textbf{Proof that the condition (4.7.) in Anton et al. (1998)
is satisfied \label{saddle point}}}

The condition is:%
\begin{equation*}
\frac{\partial ^{2}F}{\partial \overline{m}_{t}^{2}}+\delta \frac{\partial
^{2}F}{\partial \overline{m}_{t-1}^{2}}+\left\vert \frac{\partial ^{2}F}{%
\partial \overline{m}_{t}\partial \overline{m}_{t-1}}\right\vert \left(
1+\delta \right) <0\text{.}
\end{equation*}

To see this is true we first calculate the respective derivatives 
\begin{eqnarray*}
\frac{\partial ^{2}F}{\partial \overline{m}_{t}\partial \overline{m}_{t-1}}
&=&C\frac{4A_{k}^{2}\lambda }{\left( \alpha +\beta \right) ^{2}\left( \int
b_{j,t}^{\frac{1}{3}}\right) ^{6}}>0 \\
\frac{\partial ^{2}F}{\partial \overline{m}_{t}^{2}} &=&-C\frac{4A_{k}^{2}}{%
\left( \alpha +\beta \right) ^{2}\left( \int b_{j,t}^{\frac{1}{3}}\right)
^{6}}<0 \\
\frac{\partial ^{2}F}{\partial \overline{m}_{t-1}^{2}} &=&-C\frac{%
4A_{k}^{2}\lambda ^{2}}{\left( \alpha +\beta \right) ^{2}\left( \int
b_{j,t}^{\frac{1}{3}}\right) ^{6}}<0
\end{eqnarray*}%
and then substitute them into (\ref{stability}).%
\begin{eqnarray*}
&&-C\frac{4A_{k}^{2}}{\left( \alpha +\beta \right) ^{2}\left( \int b_{j,t}^{%
\frac{1}{3}}\right) ^{6}}-\delta C\frac{4A_{k}^{2}\lambda ^{2}}{\left(
\alpha +\beta \right) ^{2}\left( \int b_{j,t}^{\frac{1}{3}}\right) ^{6}}%
+\left( 1+\delta \right) C\frac{4A_{k}^{2}\lambda }{\left( \alpha +\beta
\right) ^{2}\left( \int b_{j,t}^{\frac{1}{3}}\right) ^{6}} \\
&=&-4CA_{k}^{2}\frac{\left( 1-\lambda \right) \left( 1-\lambda \delta
\right) }{\left( \alpha +\beta \right) ^{2}\left( \int b_{j,t}^{\frac{1}{3}%
}\right) ^{6}}
\end{eqnarray*}
which is clearly negative. Condition (4.7) in Anton et al. (1998) is
satisfied.

\section{\textbf{Proof Proposition \protect\ref{comparative statics meaning
and k}}}

The signs follow by simple differentiation. The only more involved cases are%
\begin{eqnarray*}
\frac{\partial \overline{m}^{\ast }}{\partial \alpha } &=&-\frac{\int b_{j}^{%
\frac{4}{3}}\left( \alpha +\beta \right) ^{2}\left( \int b_{j}^{\frac{1}{3}%
}\right) ^{5}}{C4A_{k}^{2}\left( 1-\lambda \right) }+\frac{2\left( 1-\alpha
\right) \int b_{j}^{\frac{4}{3}}\left( \alpha +\beta \right) \left( \int
b_{j}^{\frac{1}{3}}\right) ^{5}}{C4A_{k}^{2}\left( 1-\lambda \right) } \\
&&-\frac{\delta \lambda \int b_{j,t}\left( \alpha +\beta \right) ^{2}\left(
\int b_{j}^{\frac{1}{3}}\right) ^{6}}{C\left( 1-\delta \lambda \right)
4A_{k}^{2}\left( 1-\lambda \right) }+\frac{2\left( 1-\alpha \right) \delta
\lambda \int b_{j}\left( \alpha +\beta \right) \left( \int b_{j}^{\frac{1}{3}%
}\right) ^{6}}{C\left( 1-\delta \lambda \right) 4A_{k}^{2}\left( 1-\lambda
\right) } \\
&=&\left( 2\left( 1-\alpha \right) -\left( \alpha +\beta \right) \right) 
\frac{\int b_{j,t}^{\frac{4}{3}}\left( \alpha +\beta \right) \left( \int
b_{j}^{\frac{1}{3}}\right) ^{5}}{C4A_{k}^{2}\left( 1-\lambda \right) } \\
&&+\left( 2\left( 1-\alpha \right) -\left( \alpha +\beta \right) \right) 
\frac{\delta \lambda \int b_{j,t}\left( \alpha +\beta \right) \left( \int
b_{j}^{\frac{1}{3}}\right) ^{6}}{C\left( 1-\delta \lambda \right)
4A_{k}^{2}\left( 1-\lambda \right) }
\end{eqnarray*}%
\begin{equation*}
\frac{\partial \overline{m}^{\ast }}{\partial \alpha }>0\Leftrightarrow
2>3\alpha +\beta
\end{equation*}%
\begin{eqnarray*}
\frac{\partial \overline{m}^{\ast }}{\partial \lambda } &=&\frac{\left(
1-\alpha \right) \int b_{j,t}^{\frac{4}{3}}\left( \alpha +\beta \right)
^{2}\left( \int b_{j,t}^{\frac{1}{3}}\right) ^{5}}{C4A_{k}^{2}\left(
1-\lambda \right) ^{2}} \\
&&+\frac{\left( 1-\alpha \right) \delta \int b_{j,t}\left( \alpha +\beta
\right) ^{2}\left( \int b_{j,t}^{\frac{1}{3}}\right) ^{6}}{C4A_{k}^{2}}\frac{%
\left( 1-\delta \lambda \right) \left( 1-\lambda \right) -\lambda \left(
-\delta \left( 1-\lambda \right) -\left( 1-\delta \lambda \right) \right) }{%
\left( 1-\delta \lambda \right) ^{2}\left( 1-\lambda \right) ^{2}} \\
&=&\frac{\left( 1-\alpha \right) \int b_{j,t}^{\frac{4}{3}}\left( \alpha
+\beta \right) ^{2}\left( \int b_{j,t}^{\frac{1}{3}}\right) ^{5}}{%
C4A_{k}^{2}\left( 1-\lambda \right) ^{2}} \\
&&+\frac{\left( 1-\alpha \right) \delta \int b_{j,t}\left( \alpha +\beta
\right) ^{2}\left( \int b_{j,t}^{\frac{1}{3}}\right) ^{6}}{C4A_{k}^{2}}\frac{%
\left( 1-\lambda ^{2}\delta \right) }{\left( 1-\delta \lambda \right)
^{2}\left( 1-\lambda \right) ^{2}} \\
&>&0
\end{eqnarray*}%
\begin{eqnarray*}
\frac{\partial r^{\ast }}{\partial \alpha } &=&\frac{\left( -\left( \alpha
+\beta \right) +\left( 1-\alpha \right) \right) \int b_{j,t}^{\frac{4}{3}%
}\left( \int b_{j,t}^{\frac{1}{3}}\right) ^{2}}{2CA_{k}}+\frac{\left(
-\left( \alpha +\beta \right) +\left( 1-\alpha \right) \right) \delta
\lambda \int b_{j,t}\left( \int b_{j,t}^{\frac{1}{3}}\right) ^{3}}{%
2CA_{k}\left( 1-\delta \lambda \right) } \\
&=&\left( 1-\beta -2\alpha \right) \left( \frac{\int b_{j,t}^{\frac{4}{3}%
}\left( \int b_{j,t}^{\frac{1}{3}}\right) ^{2}}{2CA_{k}}+\frac{\delta
\lambda \int b_{j,t}\left( \int b_{j,t}^{\frac{1}{3}}\right) ^{3}}{%
2CA_{k}\left( 1-\delta \lambda \right) }\right) >0\Leftrightarrow 1>2\alpha
+\beta
\end{eqnarray*}%
Observe that if $\beta <1-2\alpha <2-3\alpha $ both $\frac{\partial r^{\ast }%
}{\partial \alpha }>0$ and $\frac{\partial \overline{m}^{\ast }}{\partial
\alpha }>0$ while for $1-2\alpha <\beta <2-3\alpha $ we get $\frac{\partial
r^{\ast }}{\partial \alpha }<0$ and $\frac{\partial \overline{m}^{\ast }}{%
\partial \alpha }>0$ and for $\beta >2-3\alpha $ both $\frac{\partial
r^{\ast }}{\partial \alpha }<0$ and $\frac{\partial \overline{m}^{\ast }}{%
\partial \alpha }<0$

Observe that sign of%
\begin{equation*}
sign\frac{\partial k^{\ast }}{\partial \alpha }=sign\left( \frac{3}{2}\sqrt{%
\alpha +\beta }\sqrt{1-\alpha }-\frac{1}{2}\frac{\left( \alpha +\beta
\right) ^{\frac{3}{2}}}{\sqrt{1-\alpha }}\right)
\end{equation*}%
and 
\begin{eqnarray*}
\frac{3}{2}\sqrt{\alpha +\beta }\sqrt{1-\alpha }-\frac{1}{2}\frac{\left(
\alpha +\beta \right) ^{\frac{3}{2}}}{\sqrt{1-\alpha }} &>&0 \\
\frac{3}{2}\sqrt{\alpha +\beta }\sqrt{1-\alpha } &>&\frac{1}{2}\frac{\left(
\alpha +\beta \right) ^{\frac{3}{2}}}{\sqrt{1-\alpha }} \\
3\left( 1-\alpha \right) &>&\alpha +\beta \\
3 &>&4\alpha +\beta
\end{eqnarray*}

\section{Profits in steady state and its comparative statics}

\subsection{Calculating profits in steady state \label{profit final}}

Substituting $m^{\ast },r^{\ast },$ $k^{\ast }$ and $e^{\ast }$ into profits
we get:

{\footnotesize
\begin{eqnarray}
\Pi ^{\ast } &=&\frac{\left( 1-\alpha \right) }{1-\delta }\left( \int
b_{j}^{2}\frac{\alpha }{A_{e}}\right)  \label{pi*} \\
&&+\frac{\left( 1-\alpha \right) }{1-\delta }\left( \frac{\left( 1-\alpha
\right) \left( \alpha +\beta \right) ^{2}\left( \int b_{j}^{\frac{1}{3}%
}\right) ^{4}\left( \int b_{j}^{\frac{4}{3}}\right) ^{2}}{4CA_{k}^{2}}+\frac{%
\left( 1-\alpha \right) \delta \lambda \left( \alpha +\beta \right)
^{2}\left( \int b_{j}^{\frac{1}{3}}\right) ^{5}\int b_{j}\left( \int b_{j}^{%
\frac{4}{3}}\right) }{4CA_{k}^{2}\left( 1-\delta \lambda \right) }\right) 
\notag \\
&&+\frac{\left( 1-\alpha \right) }{1-\delta }\left( \left( \frac{\left(
1-\alpha \right) \lambda \left( \alpha +\beta \right) ^{2}\left( \int b_{j}^{%
\frac{1}{3}}\right) ^{5}\int b_{j}\int b_{j}^{\frac{4}{3}}}{%
C4A_{k}^{2}\left( 1-\lambda \right) }+\frac{\left( 1-\alpha \right) \delta
\lambda ^{2}\left( \alpha +\beta \right) ^{2}\left( \int b_{j}^{\frac{1}{3}%
}\right) ^{6}\left( \int b_{j}\right) ^{2}}{C\left( 1-\delta \lambda \right)
4A_{k}^{2}\left( 1-\lambda \right) }\right) \right)  \notag \\
&&-\frac{1}{1-\delta }\frac{C}{2}\left( \frac{\left( 1-\alpha \right) \left(
\alpha +\beta \right) \left( \int b_{j}^{\frac{1}{3}}\right) ^{2}\int b_{j}^{%
\frac{4}{3}}}{2CA_{k}}+\frac{\left( 1-\alpha \right) \delta \lambda \left(
\alpha +\beta \right) \left( \int b_{j}^{\frac{1}{3}}\right) ^{3}\int b_{j}}{%
2CA_{k}\left( 1-\delta \lambda \right) }\right) ^{2}  \notag
\end{eqnarray}
}
Combining lines 2 and 3 above%
{\footnotesize
\begin{equation*}
\frac{\left( 1-\alpha \right) ^{2}\left( \alpha +\beta \right) ^{2}}{\left(
1-\delta \right) 4CA_{k}^{2}}\left( \left( \int b_{j}^{\frac{1}{3}}\right)
^{4}\left( \int b_{j}^{\frac{4}{3}}\right) ^{2}+\frac{\lambda \left(
1-2\delta \lambda +\delta \right) }{\left( 1-\lambda \right) \left( 1-\delta
\lambda \right) }\left( \int b_{j}^{\frac{1}{3}}\right) ^{5}\int b_{j}\int
b_{j}^{\frac{4}{3}}+\frac{\delta \lambda ^{2}\left( \int b_{j}^{\frac{1}{3}%
}\right) ^{6}\left( \int b_{j}\right) ^{2}}{\left( 1-\delta \lambda \right)
\left( 1-\lambda \right) }\right)
\end{equation*}%
}
so that 
\begin{eqnarray*}
\Pi ^{\ast } &=&\frac{\left( 1-\alpha \right) }{1-\delta }\left( \int
b_{j}^{2}\frac{\alpha }{A_{e}}\right) \\
&&+\frac{\left( 1-\alpha \right) ^{2}\left( \alpha +\beta \right) ^{2}}{%
\left( 1-\delta \right) 4CA_{k}^{2}}\left( 
\begin{array}{c}
\left( \int b_{j}^{\frac{1}{3}}\right) ^{4}\left( \int b_{j}^{\frac{4}{3}%
}\right) ^{2}+\frac{\lambda \left( 1-2\delta \lambda +\delta \right) }{%
\left( 1-\lambda \right) \left( 1-\delta \lambda \right) }\left( \int b_{j}^{%
\frac{1}{3}}\right) ^{5}\int b_{j}\int b_{j}^{\frac{4}{3}} \\ 
+\frac{\delta \lambda ^{2}\left( \int b_{j}^{\frac{1}{3}}\right) ^{6}\left(
\int b_{j}\right) ^{2}}{\left( 1-\delta \lambda \right) \left( 1-\lambda
\right) }%
\end{array}%
\right) \\
&&-\frac{\left( 1-\alpha \right) ^{2}\left( \alpha +\beta \right) ^{2}}{%
\left( 1-\delta \right) 8A_{k}^{2}C}\left( \left( \int b_{j}^{\frac{1}{3}%
}\right) ^{2}\int b_{j}^{\frac{4}{3}}+\frac{\delta \lambda }{\left( 1-\delta
\lambda \right) }\left( \int b_{j}^{\frac{1}{3}}\right) ^{3}\int
b_{j}\right) ^{2}
\end{eqnarray*}%
Calculating the square in the third line we get

\begin{eqnarray*}
\Pi ^{\ast } &=&\frac{\left( 1-\alpha \right) }{1-\delta }\left( \int
b_{j}^{2}\frac{\alpha }{A_{e}}\right) \\
&&+\frac{\left( 1-\alpha \right) ^{2}\left( \alpha +\beta \right) ^{2}}{%
\left( 1-\delta \right) 4CA_{k}^{2}}\left( 
\begin{array}{c}
\left( \int b_{j}^{\frac{1}{3}}\right) ^{4}\left( \int b_{j}^{\frac{4}{3}%
}\right) ^{2}+\frac{\lambda \left( 1-2\delta \lambda +\delta \right) }{%
\left( 1-\lambda \right) \left( 1-\delta \lambda \right) }\left( \int b_{j}^{%
\frac{1}{3}}\right) ^{5}\int b_{j}\int b_{j}^{\frac{4}{3}} \\ 
+\frac{\delta \lambda ^{2}\left( \int b_{j}^{\frac{1}{3}}\right) ^{6}\left(
\int b_{j}\right) ^{2}}{\left( 1-\delta \lambda \right) \left( 1-\lambda
\right) }%
\end{array}%
\right) \\
&&-\frac{\left( 1-\alpha \right) ^{2}\left( \alpha +\beta \right) ^{2}}{%
\left( 1-\delta \right) 8A_{k}^{2}C}\left( 
\begin{array}{c}
\left( \int b_{j}^{\frac{1}{3}}\right) ^{4}\left( \int b_{j}^{\frac{4}{3}%
}\right) ^{2}+\frac{2\delta \lambda }{\left( 1-\delta \lambda \right) }%
\left( \int b_{j}^{\frac{1}{3}}\right) ^{5}\int b_{j}\int b_{j}^{\frac{4}{3}}
\\ 
+\frac{\delta ^{2}\lambda ^{2}}{2\left( 1-\delta \lambda \right) ^{2}}\left(
\int b_{j}^{\frac{1}{3}}\right) ^{6}\left( \int b_{j}\right) ^{2}%
\end{array}%
\right)
\end{eqnarray*}%
Finally, collecting terms this simplifies to

\begin{eqnarray*}
\Pi ^{\ast } &=&\frac{\left( 1-\alpha \right) }{1-\delta }\frac{\alpha }{%
A_{e}}\left( \int b_{j}^{2}\right) +\frac{\left( 1-\alpha \right) ^{2}\left(
\alpha +\beta \right) ^{2}}{\left( 1-\delta \right) 4CA_{k}^{2}} \\
&&\times \left( 
\begin{array}{c}
\frac{1}{2}\left( \int b_{j}^{\frac{1}{3}}\right) ^{4}\left( \int b_{j}^{%
\frac{4}{3}}\right) ^{2}+\frac{\lambda }{\left( 1-\lambda \right) }\left(
\int b_{j}^{\frac{1}{3}}\right) ^{5}\int b_{j}\int b_{j}^{\frac{4}{3}} \\ 
+\frac{\delta \lambda ^{2}\left( 4-3\delta \lambda -\delta \right) }{4\left(
1-\delta \lambda \right) ^{2}\left( 1-\lambda \right) }\left( \int b_{j}^{%
\frac{1}{3}}\right) ^{6}\left( \int b_{j}\right) ^{2}%
\end{array}%
\right)
\end{eqnarray*}

\subsection{Proof of Comparative statics with respect to steady state profits%
}

\textbf{Proof of Proposition \ref{comparative static pi*}}

The fact that $\Pi ^{\ast }$ is decreasing in $A_{e}$ and $A_{k}$, and $C$
is obvious by inspection. So is the fact that $\Pi ^{\ast }$ is increasing
in $\beta .$

Concerning the effects of $\lambda $ and $\delta $ on $\Pi ^{\ast }$ the
only place where there is a possible ambiguity is in $\frac{\delta \lambda
^{2}\left( 4-3\delta \lambda -\delta \right) }{4\left( 1-\delta \lambda
\right) ^{2}\left( 1-\lambda \right) }.$ Let 
\begin{equation*}
G\left( \delta ,\lambda \right) =\frac{\lambda ^{2}\delta \left( 4-\delta
-3\lambda \delta \right) }{2\left( 1-\delta \lambda \right) ^{2}\left(
1-\lambda \right) }
\end{equation*}

\begin{equation*}
\frac{\partial G}{\partial \delta }=\frac{\lambda ^{2}\left( 2-\delta
(1+\lambda )\right) }{(1-\lambda )\,(1-\delta \lambda )^{3}}\geq 0
\end{equation*}

Since $\delta $ and $\lambda $ are in $[0,1]$. Now, 
\begin{equation*}
\frac{\partial G}{\partial \lambda }=\frac{\delta \lambda \;P(\delta
,\lambda )}{(1-\lambda )^{2}\,(1-\delta \lambda )^{3}},
\end{equation*}%
where 
\begin{equation*}
P(\delta ,\lambda )=2\delta ^{2}\lambda ^{2}+\delta \lambda ^{2}-4\delta
\lambda -\delta -2\lambda +4.
\end{equation*}

The sign of $\frac{\partial G}{\partial \lambda }$ is the sign of $P(\delta
,\lambda )$.

For fixed $\delta \in (0,1)$, $P$ is a quadratic polynomial in $\lambda $: 
\begin{equation*}
P(\lambda )=(2\delta ^{2}+\delta )\lambda ^{2}-(4\delta +2)\lambda
+(4-\delta ),
\end{equation*}%
with leading coefficient $2\delta ^{2}+\delta >0$. Its discriminant is 
\begin{equation*}
\Delta =(4\delta +2)^{2}-4(2\delta ^{2}+\delta )(4-\delta )=4(\delta
-1)^{2}(2\delta +1)>0,\quad (0<\delta <1),
\end{equation*}%
so there are two real roots. They can be written as 
\begin{equation*}
\lambda _{\pm }=\frac{1}{\delta }\pm \frac{1-\delta }{\delta \sqrt{2\delta +1%
}}.
\end{equation*}

We now show that both roots exceed $1$.

First, clearly 
\begin{equation*}
\lambda _{+}=\frac{1}{\delta }+\frac{1-\delta }{\delta \sqrt{2\delta +1}}%
\geq 1.
\end{equation*}

For $\lambda _{-}$, observe 
\begin{equation*}
\lambda _{-}>1\iff \frac{1}{\delta }-\frac{1-\delta }{\delta \sqrt{2\delta +1%
}}>1\iff 1-\frac{1-\delta }{\sqrt{2\delta +1}}>\delta .
\end{equation*}%
Rearranging gives 
\begin{equation*}
1-\delta >\frac{1-\delta }{\sqrt{2\delta +1}}.
\end{equation*}%
which is true on our domain. Hence $\lambda _{-}>1$. Therefore 
\begin{equation*}
\lambda _{-}>1,\qquad \lambda _{+}>1.
\end{equation*}

Since $P(\delta ,0)=4-\delta >0,$ and $P\left( \delta ,\lambda \right) $ is
quadratic in $\lambda ,$ it follows that $P(\delta ,\lambda )>0$ for all $%
0<\lambda <1$ as we just showed both roots are bigger than one. Thus, 
\begin{equation*}
\frac{\partial G}{\partial \lambda }>0.
\end{equation*}

Finally%
\begin{eqnarray*}
\Pi ^{\ast } &=&\frac{\left( 1-\alpha \right) }{1-\delta }\frac{\alpha }{%
A_{e}}\left( \int b_{j}^{2}\right) +\frac{\left( 1-\alpha \right) ^{2}\left(
\alpha +\beta \right) ^{2}}{\left( 1-\delta \right) 4CA_{k}^{2}} \\
&&\times \left( \frac{1}{2}\left( \int b_{j}^{\frac{1}{3}}\right) ^{4}\left(
\int b_{j}^{\frac{4}{3}}\right) ^{2}+\frac{\lambda }{\left( 1-\lambda
\right) }\left( \int b_{j}^{\frac{1}{3}}\right) ^{5}\int b_{j}\int b_{j}^{%
\frac{4}{3}}\right. \\
&&+\left. \frac{\delta \lambda ^{2}\left( 4-3\delta \lambda -\delta \right) 
}{4\left( 1-\delta \lambda \right) ^{2}\left( 1-\lambda \right) }\left( \int
b_{j}^{\frac{1}{3}}\right) ^{6}\left( \int b_{j}\right) ^{2}\right)
\end{eqnarray*}%
$\allowbreak $%
\begin{eqnarray*}
\frac{\partial \Pi ^{\ast }}{\partial \alpha } &=&\frac{\left( 1-2\alpha
\right) }{1-\delta }\frac{1}{A_{e}}\left( \int b_{j}^{2}\right) +\frac{%
2\left( 1-\alpha \right) \left( \alpha +\beta \right) \left( \left( 1-\alpha
\right) -\left( \alpha +\beta \right) \right) }{\left( 1-\delta \right)
4CA_{k}^{2}} \\
&&\times \left( \frac{1}{2}\left( \int b_{j}^{\frac{1}{3}}\right) ^{4}\left(
\int b_{j}^{\frac{4}{3}}\right) ^{2}+\frac{\lambda }{\left( 1-\lambda
\right) }\left( \int b_{j}^{\frac{1}{3}}\right) ^{5}\int b_{j}\int b_{j}^{%
\frac{4}{3}}\right. \\
&&+\left. \frac{\delta \lambda ^{2}\left( 4-3\delta \lambda -\delta \right) 
}{4\left( 1-\delta \lambda \right) ^{2}\left( 1-\lambda \right) }\left( \int
b_{j}^{\frac{1}{3}}\right) ^{6}\left( \int b_{j}\right) ^{2}\right) \\
&=&\frac{\left( 1-2\alpha \right) }{1-\delta }\frac{1}{A_{e}}\left( \int
b_{j}^{2}\right) +\frac{2\left( 1-\alpha \right) \left( \alpha +\beta
\right) \left( 1-2\alpha -\beta \right) }{\left( 1-\delta \right) 4CA_{k}^{2}%
} \\
&&\times \left( \frac{1}{2}\left( \int b_{j}^{\frac{1}{3}}\right) ^{4}\left(
\int b_{j}^{\frac{4}{3}}\right) ^{2}+\frac{\lambda }{\left( 1-\lambda
\right) }\left( \int b_{j}^{\frac{1}{3}}\right) ^{5}\int b_{j}\int b_{j}^{%
\frac{4}{3}}\right. \\
&&+\left. \frac{\delta \lambda ^{2}\left( 4-3\delta \lambda -\delta \right) 
}{4\left( 1-\delta \lambda \right) ^{2}\left( 1-\lambda \right) }\left( \int
b_{j}^{\frac{1}{3}}\right) ^{6}\left( \int b_{j}\right) ^{2}\right)
\end{eqnarray*}

and the result follows.

\section{\textbf{Examples for comparative statics different talent
distributions\label{examples lognormal} }}

Assume that $b^{p}$ is lognormally distributed, then the formula for the $%
n^{th}$ moment of the lognormal random variable is given by 
\begin{equation}
\int \left( b^{p}\right) ^{n}=\exp \left( n\mu +\frac{1}{2}\sigma
^{2}n^{2}\right).  \label{moments log normal}
\end{equation}%
where $\mu $ refers to the mean and $\sigma ^{2}$ to the variance of the
underlying normal distribution of the lognormal distribution and $n$ is a
real number.

Let us start with the baseline variable being $b,$which is an intermediate
case not covered in proposition \ref{SD} since $p=1$. We first look at
optimal meaning $\overline{m}^{\ast }$ given by (\ref{steady state m}) as

\begin{equation*}
\overline{m}^{\ast }=\frac{\left( 1-\alpha \right) \int b_{j}^{\frac{4}{3}%
}\left( \alpha +\beta \right) ^{2}\left( \int b_{j}^{\frac{1}{3}}\right) ^{5}%
}{C4A_{k}^{2}\left( 1-\lambda \right) }+\frac{\left( 1-\alpha \right) \delta
\lambda \int b_{j}\left( \alpha +\beta \right) ^{2}\left( \int b_{j}^{\frac{1%
}{3}}\right) ^{6}}{C\left( 1-\delta \lambda \right) 4A_{k}^{2}\left(
1-\lambda \right) }
\end{equation*}

Note that $\int b_{j}^{\frac{1}{3}}=\exp \left( \frac{1}{3}\mu +\frac{1}{18}%
\sigma ^{2}\right) $, $\int b_{j}=\exp \left( \mu +\frac{1}{2}\sigma
^{2}\right) ,$and $\int b_{j}^{\frac{4}{3}}=\exp \left( \frac{4}{3}\mu +%
\frac{16}{18}\sigma ^{2}\right)$. Thus

\begin{eqnarray}
\overline{m}^{\ast } &=&\frac{\left( 1-\alpha \right) \exp \left( \frac{4}{3}%
\mu +\frac{16}{18}\sigma ^{2}\right) \left( \alpha +\beta \right) ^{2}\left(
\exp \left( \frac{1}{3}\mu +\frac{1}{18}\sigma ^{2}\right) \right) ^{5}}{%
C4A_{k}^{2}\left( 1-\lambda \right) }  \label{m* lognormal}
 \\
&&+\frac{\left( 1-\alpha \right) \delta \lambda \exp \left( \mu +\frac{1}{2}%
\sigma ^{2}\right) \left( \alpha +\beta \right) ^{2}\left( \exp \left( \frac{%
1}{3}\mu +\frac{1}{18}\sigma ^{2}\right) \right) ^{6}}{C\left( 1-\delta
\lambda \right) 4A_{k}^{2}\left( 1-\lambda \right) }  \notag \\
&=&\frac{\left( 1-\alpha \right) \exp \left( 3\mu +\frac{7}{6}\sigma
^{2}\right) \left( \alpha +\beta \right) ^{2}}{C4A_{k}^{2}\left( 1-\lambda
\right) }+\frac{\left( 1-\alpha \right) \delta \lambda \exp \left( 3\mu +%
\frac{5}{6}\sigma ^{2}\right) \left( \alpha +\beta \right) ^{2}}{C\left(
1-\delta \lambda \right) 4A_{k}^{2}\left( 1-\lambda \right) } \notag
\end{eqnarray}

We now compare two talent distributions that can be ranked by SOSD. In
particular, we examine how steady state meaning is affected by a change in
variance without changing the mean. We use $\sigma ^{2}+\Gamma $ which
increases variance. But to keep the mean constant we use $\mu -\frac{1}{2}%
\Gamma ,$ so the mean $\mu -\frac{1}{2}\Gamma +\frac{1}{2}\sigma ^{2}+\frac{1%
}{2}\Gamma $ is constant. This increase in dispersion affects steady state
meaning as follows:

{\footnotesize
\begin{eqnarray}
\overline{m}^{\ast }\left( \Gamma \right) &=&\frac{\left( 1-\alpha \right)
\exp \left( 3\mu -\frac{3}{2}\Gamma +\frac{7}{6}\sigma ^{2}+\frac{7}{6}%
\Gamma \right) \left( \alpha +\beta \right) ^{2}}{C4A_{k}^{2}\left(
1-\lambda \right) }   \label{m* Gamma lognormal} \\
&&+\frac{\left( 1-\alpha \right) \delta \lambda \exp \left( 3\mu -\frac{3}{2}%
\Gamma +\frac{5}{6}\sigma ^{2}+\frac{5}{6}\Gamma \right) \left( \alpha
+\beta \right) ^{2}}{C\left( 1-\delta \lambda \right) 4A_{k}^{2}\left(
1-\lambda \right) }  \notag \\
&=&\frac{\left( 1-\alpha \right) \exp \left( 3\mu +\frac{7}{6}\sigma ^{2}-%
\frac{2}{6}\Gamma \right) \left( \alpha +\beta \right) ^{2}}{%
C4A_{k}^{2}\left( 1-\lambda \right) }+\frac{\left( 1-\alpha \right) \delta
\lambda \exp \left( 3\mu +\frac{5}{6}\sigma ^{2}-\frac{4}{6}\Gamma \right)
\left( \alpha +\beta \right) ^{2}}{C\left( 1-\delta \lambda \right)
4A_{k}^{2}\left( 1-\lambda \right) } \notag
\end{eqnarray}%
}
and it is clear from (\ref{m* lognormal}) and (\ref{m* Gamma lognormal})
that 
\begin{equation*}
\overline{m}^{\ast }\left( \Gamma \right) -\overline{m}^{\ast }<0
\end{equation*}%
So an increase in the variance that keeps the mean constant decreases $%
m^{\ast }$ when $\ b_{i}$ is lognormally distributed.

Now we consider what happens to $r^{\ast }$ in this case. Under the
log-normal assumption

\begin{eqnarray}
r^{\ast } &=&\frac{\left( 1-\alpha \right) \int b_{j,t}^{\frac{4}{3}}\left(
\alpha +\beta \right) \left( \int b_{j,t}^{\frac{1}{3}}\right) ^{2}}{2CA_{k}}%
+\frac{\left( 1-\alpha \right) \delta \lambda \int b_{j,t}\left( \alpha
+\beta \right) \left( \int b_{j,t}^{\frac{1}{3}}\right) ^{3}}{2CA_{k}\left(
1-\delta \lambda \right) }   \label{r lognormal} \\
&=&\frac{\left( 1-\alpha \right) \exp \left( \frac{4}{3}\mu +\frac{16}{18}%
\sigma ^{2}\right) \left( \alpha +\beta \right) \left( \exp \left( \frac{1}{3%
}\mu +\frac{1}{18}\sigma ^{2}\right) \right) ^{2}}{2CA_{k}}  \notag \\
&&+\frac{\left( 1-\alpha \right) \delta \lambda \exp \left( \mu +\frac{1}{2}%
\sigma ^{2}\right) \left( \alpha +\beta \right) \left( \exp \left( \frac{1}{3%
}\mu +\frac{1}{18}\sigma ^{2}\right) \right) ^{3}}{2CA_{k}\left( 1-\delta
\lambda \right) }  \notag \\
&=&\frac{\left( 1-\alpha \right) \exp \left( \frac{6}{3}\mu +\frac{18}{18}%
\sigma ^{2}\right) \left( \alpha +\beta \right) }{2CA_{k}}+\frac{\left(
1-\alpha \right) \delta \lambda \exp \left( \frac{6}{3}\mu +\frac{12}{18}%
\sigma ^{2}\right) \left( \alpha +\beta \right) }{2CA_{k}\left( 1-\delta
\lambda \right) } \notag
\end{eqnarray}%
Introducing the increase in dispersion we get%
{\footnotesize
\begin{eqnarray}
r^{\ast }\left( \Gamma \right) &=&\frac{\left( 1-\alpha \right) \exp \left( 
\frac{6}{3}\left( \mu -\frac{1}{2}\Gamma \right) +\frac{18}{18}\left( \sigma
^{2}+\Gamma \right) \right) \left( \alpha +\beta \right) }{2CA_{k}} \label{r gamma lognormal}
\\
&&+\frac{\left( 1-\alpha \right) \delta \lambda \exp \left( \frac{6}{3}%
\left( \mu -\frac{1}{2}\Gamma \right) +\frac{12}{18}\left( \sigma
^{2}+\Gamma \right) \right) \left( \alpha +\beta \right) }{2CA_{k}\left(
1-\delta \lambda \right) }  \notag \\
&=&\frac{\left( 1-\alpha \right) \exp \left( \frac{6}{3}\mu +\frac{18}{18}%
\sigma ^{2}\right) \left( \alpha +\beta \right) }{2CA_{k}}+\frac{\left(
1-\alpha \right) \delta \lambda \exp \left( \frac{6}{3}\mu +\frac{12}{18}%
\sigma ^{2}-\frac{6}{18}\Gamma \right) \left( \alpha +\beta \right) }{%
2CA_{k}\left( 1-\delta \lambda \right) }  \notag
\end{eqnarray}%
}
and it is clear from (\ref{r lognormal}) and (\ref{r gamma lognormal}) that 
\begin{equation*}
r^{\ast }\left( \Gamma \right) -r^{\ast }<0
\end{equation*}

So an increase in the variance that keeps the mean constant decreases $%
r^{\ast }$ when $\ b_{i}$ is lognormally distributed$.$

We now study $k^{\ast }$ given by (\ref{steady state k}) which under the
assumption of our distribution becomes 
\begin{eqnarray}
k^{\ast } &=&\frac{\left( 1-\alpha \right) ^{\frac{1}{2}}\left( \alpha
+\beta \right) ^{\frac{3}{2}}}{2A_{k}}\exp \left( \frac{1}{3}\mu +\frac{1}{18%
}\sigma ^{2}\right) \left( \exp \left( \frac{2}{3}\mu +\frac{4}{18}\sigma
^{2}\right) \right)  \label{k* lognormal} \\
&&\left( \frac{\exp \left( \frac{4}{3}\mu +\frac{16}{18}\sigma ^{2}\right)
\left( \exp \left( \frac{1}{3}\mu +\frac{1}{18}\sigma ^{2}\right) \right)
^{2}}{2CA_{k}}+\frac{\delta \lambda \exp \left( \mu +\frac{1}{2}\sigma
^{2}\right) \left( \exp \left( \frac{1}{3}\mu +\frac{1}{18}\sigma
^{2}\right) \right) ^{3}}{2CA_{k}\left( 1-\delta \lambda \right) }\right) ^{%
\frac{1}{2}}  \notag \\
&=&\frac{\left( 1-\alpha \right) ^{\frac{1}{2}}\left( \alpha +\beta \right)
^{\frac{3}{2}}}{2A_{k}}\exp \left( \mu +\frac{5}{18}\sigma ^{2}\right)
\left( \frac{\exp \left( 2\mu +\sigma ^{2}\right) }{2CA_{k}}+\frac{\delta
\lambda \exp \left( 2\mu +\frac{12}{18}\sigma ^{2}\right) }{2CA_{k}\left(
1-\delta \lambda \right) }\right) ^{\frac{1}{2}} \notag
\end{eqnarray}%
Introducing the increase in dispersion we get

{\footnotesize
\begin{eqnarray}
k^{\ast }\left( \Gamma \right) &=&\frac{\left( 1-\alpha \right) ^{\frac{1}{2}%
}\left( \alpha +\beta \right) ^{\frac{3}{2}}}{2A_{k}}\exp \left( \left( \mu -%
\frac{1}{2}\Gamma \right) +\frac{5}{18}\left( \sigma ^{2}+\Gamma \right)
\right)  \label{k* lognormal Gamma} \\
&&\left( \frac{\exp \left( 2\left( \mu -\frac{1}{2}\Gamma \right) +\left(
\sigma ^{2}+\Gamma \right) \right) }{2CA_{k}}+\frac{\delta \lambda \exp
\left( 2\left( \mu -\frac{1}{2}\Gamma \right) +\frac{12}{18}\left( \sigma
^{2}+\Gamma \right) \right) }{2CA_{k}\left( 1-\delta \lambda \right) }%
\right) ^{\frac{1}{2}}  \notag \\
&=&\frac{\left( 1-\alpha \right) ^{\frac{1}{2}}\left( \alpha +\beta \right)
^{\frac{3}{2}}}{2A_{k}}\exp \left( \mu +\frac{5}{18}\sigma ^{2}-\frac{4}{18}%
\Gamma \right) \left( \frac{\exp \left( 2\mu +\sigma ^{2}\right) }{2CA_{k}}+%
\frac{\delta \lambda \exp \left( 2\mu +\frac{12}{18}\sigma ^{2}-\frac{6}{18}%
\Gamma \right) }{2CA_{k}\left( 1-\delta \lambda \right) }\right) ^{\frac{1}{2%
}}  \notag
\end{eqnarray}%
}
and it is clear from (\ref{k* lognormal}) and (\ref{k* lognormal Gamma})
that 
\begin{equation*}
k^{\ast }\left( \Gamma \right) -k^{\ast }<0.
\end{equation*}%
So an increase in the variance that keeps the mean constant decreases $%
k^{\ast \ast }$ when $\ b_{i}$ is lognormally distributed.

Introducing the lognormal assumption into steady state utility given by (\ref%
{steady state utility}) we get

\begin{eqnarray*}
u^{\ast } &=&b_{i,t}^{\frac{4}{3}}\left( \left( \frac{3\left( \alpha +\beta
\right) ^{2}}{8A_{k}}\left( \int b_{j}^{\frac{1}{3}}\right) ^{2}\right)
r^{\ast }+\lambda \overline{m}^{\ast }\,\right) +\frac{b_{i,t}^{2}}{2}\frac{%
\alpha ^{2}}{A_{e}} \\
&=&b_{i,t}^{\frac{4}{3}}\left( \left( \frac{3\left( \alpha +\beta \right)
^{2}}{8A_{k}}\left( \exp \left( \frac{1}{3}\mu +\frac{1}{18}\sigma
^{2}\right) \right) ^{2}\right) r^{\ast }+\lambda \overline{m}^{\ast
}\,\right) +\frac{b_{i,t}^{2}}{2}\frac{\alpha ^{2}}{A_{e}}
\end{eqnarray*}

and with the increase in dispersion%
{\footnotesize
\begin{eqnarray*}
u^{\ast }\left( \Gamma \right) &=&b_{i,t}^{\frac{4}{3}}\left( \left( \frac{%
3\left( \alpha +\beta \right) ^{2}}{8A_{k}}\left( \exp \left( \frac{1}{3}%
\left( \mu -\frac{1}{2}\Gamma \right) +\frac{1}{18}\left( \sigma ^{2}+\Gamma
\right) \right) \right) ^{2}\right) r^{\ast }\left( \Gamma \right) +\lambda 
\overline{m}^{\ast }\,\left( \Gamma \right) \right) +\frac{b_{i,t}^{2}}{2}%
\frac{\alpha ^{2}}{A_{e}} \\
&=&b_{i,t}^{\frac{4}{3}}\left( \left( \frac{3\left( \alpha +\beta \right)
^{2}}{8A_{k}}\left( \exp \left( \frac{1}{3}\mu +\frac{1}{18}\sigma ^{2}-%
\frac{2}{18}\Gamma \right) \right) ^{2}\right) r^{\ast }\left( \Gamma
\right) +\lambda \overline{m}^{\ast }\,\left( \Gamma \right) \right) +\frac{%
b_{i,t}^{2}}{2}\frac{\alpha ^{2}}{A_{e}}
\end{eqnarray*}
}
Since $\overline{m}^{\ast }\,\left( \Gamma \right) -\overline{m}^{\ast }<0$,
$r^{\ast }\left( \Gamma \right) -r^{\ast }<0$, and $\exp \left( \frac{1}{3%
}\mu +\frac{1}{18}\sigma ^{2}-\frac{2}{18}\Gamma \right) -\exp \left( \frac{1%
}{3}\mu +\frac{1}{18}\sigma ^{2}\right) <0$, we have that $u^{\ast }\left(
\Gamma \right) -u^{\ast }<0.$

Finally, we study steady state profits given by (\ref{steady state profits})
when $b_{i}$ is lognormally distributed by introducing the corresponding
expressions of (\ref{moments log normal}) and simplifying we get

{\footnotesize
\begin{eqnarray}
\Pi ^{\ast } &=&\frac{1}{1-\delta }\left( 1-\alpha \right) \left( \exp
\left( 2\mu +2\sigma ^{2}\right) \frac{\alpha }{A_{e}}\right) \label{Pi* lognormal} \\
&&+\frac{1}{1-\delta }\left( 1-\alpha \right) ^{2}\frac{\left( \alpha +\beta
\right) ^{2}}{4CA_{k}^{2}}\ast   \notag \\
&&\left( \frac{1}{2}\exp \left( 4\mu +2\sigma ^{2}\right) +\frac{\lambda }{%
1-\lambda }\exp \left( 4\mu +\frac{30}{18}\sigma ^{2}\right) +\frac{\lambda
^{2}\delta \left( 2-\delta -\delta \lambda \right) }{2\left( 1-\lambda
\right) \left( 1-\delta \lambda \right) ^{2}}\left( \exp \left( 4\mu +\frac{4%
}{3}\sigma ^{2}\right) \right) \right) \notag
\end{eqnarray}
}
while with the increase in dispersion we get 
\begin{eqnarray}
\Pi ^{\ast }\left( \Gamma \right) &=&\frac{\left( 1-\alpha \right) }{%
1-\delta }\left( \exp \left( 2\mu +2\sigma ^{2}+\Gamma \right) \frac{\alpha 
}{A_{e}}\right)  \label{Pi* Lognormal Gamma} \\
&&+\frac{\left( 1-\alpha \right) ^{2}}{1-\delta }\frac{\left( \alpha +\beta
\right) ^{2}}{4CA_{k}^{2}}\ast  \notag \\
&&\left( \frac{1}{2}\exp \left( 4\mu +2\sigma ^{2}\right) +\frac{\lambda }{%
1-\lambda }\exp \left( 4\mu +\frac{30}{18}\sigma ^{2}-\frac{6}{18}\Gamma
\right) \right.  \notag \\
&&\left. +\frac{\lambda ^{2}\delta \left( 2-\delta -\delta \lambda \right) }{%
2\left( 1-\lambda \right) \left( 1-\delta \lambda \right) ^{2}}\left( \exp
\left( 4\mu +\frac{4}{3}\sigma ^{2}-\frac{2}{3}\Gamma \right) \right) \right)
\notag
\end{eqnarray}

Comparing $\Pi ^{\ast }\left( \Gamma \right) $ to $\Pi ^{\ast }$ we can see
that the first line in (\ref{Pi* Lognormal Gamma}) is bigger than the first
line in (\ref{Pi* lognormal}) but the opposite hold for the third line in (%
\ref{Pi* Lognormal Gamma}) versus the third line in (\ref{Pi* lognormal})
(which are both multiplied by the same second line). If $A_{e}$ is much
smaller than $A_{k}$ we can get that 
\begin{equation*}
\Pi ^{\ast }\left( \Gamma \right) -\Pi ^{\ast }>0
\end{equation*}%
but the opposite holds otherwise. This clearly illustrates that the cases
not covered by proposition \ref{SD} have to be studied in detail and a SOSD
shift (which is a reduction in dispersion keeping the mean constant) might
increase or decrease the studied steady state outcome.

To illustrate the case of a convex baseline distribution function which is
covered in proposition \ref{SD} assume that $b^{\frac{1}{3}}$ is lognormally
distributed (hence $p=\frac{1}{3})$ and let's calculate how an increase in
dispersion affects $\overline{m}^{\ast }$ given by

\begin{equation*}
\overline{m}^{\ast }=\frac{\left( 1-\alpha \right) \int b_{j}^{\frac{4}{3}%
}\left( \alpha +\beta \right) ^{2}\left( \int b_{j}^{\frac{1}{3}}\right) ^{5}%
}{C4A_{k}^{2}\left( 1-\lambda \right) }+\frac{\left( 1-\alpha \right) \delta
\lambda \int b_{j}\left( \alpha +\beta \right) ^{2}\left( \int b_{j}^{\frac{1%
}{3}}\right) ^{6}}{C\left( 1-\delta \lambda \right) 4A_{k}^{2}\left(
1-\lambda \right) }
\end{equation*}

Note then that $\int b_{j}^{\frac{1}{3}}=\exp \left( \mu +\frac{1}{2}\sigma
^{2}\right)$, $\int b_{j}=\exp \left( 3\mu +\frac{9}{2}\sigma ^{2}\right)
,$and $\int b_{j}^{\frac{4}{3}}=\exp \left( 4\mu +\frac{16}{2}\sigma
^{2}\right)$. So

\begin{eqnarray}
\overline{m}^{\ast } &=&\frac{\left( 1-\alpha \right) \exp \left( 4\mu +%
\frac{16}{2}\sigma ^{2}\right) \left( \alpha +\beta \right) ^{2}\left( \exp
\left( \mu +\frac{1}{2}\sigma ^{2}\right) \right) ^{5}}{C4A_{k}^{2}\left(
1-\lambda \right) }  \label{m* lognormal 1/3} \\
&&+\frac{\left( 1-\alpha \right) \delta \lambda \exp \left( 3\mu +\frac{9}{2}%
\sigma ^{2}\right) \left( \alpha +\beta \right) ^{2}\left( \exp \left( \mu +%
\frac{1}{2}\sigma ^{2}\right) \right) ^{6}}{C\left( 1-\delta \lambda \right)
4A_{k}^{2}\left( 1-\lambda \right) }  \notag \\
&&\frac{\left( 1-\alpha \right) \exp \left( 9\mu +\frac{21}{2}\sigma
^{2}\right) \left( \alpha +\beta \right) ^{2}}{C4A_{k}^{2}\left( 1-\lambda
\right) }+\frac{\left( 1-\alpha \right) \delta \lambda \exp \left( 9\mu +%
\frac{15}{2}\sigma ^{2}\right) \left( \alpha +\beta \right) ^{2}}{C\left(
1-\delta \lambda \right) 4A_{k}^{2}\left( 1-\lambda \right) }
\notag
\end{eqnarray}

Change $\sigma ^{2}+\Gamma $ which increases variance but keep mean constant
by decreasing $\mu -\frac{1}{2}\Gamma ,$ so $\left( \mu -\frac{1}{2}\Gamma +%
\frac{1}{2}\sigma ^{2}+\frac{1}{2}\Gamma \right) $ is constant

{\footnotesize
\begin{eqnarray}
\overline{m}^{\ast }\left( \Gamma \right) &=&\frac{\left( 1-\alpha \right)
\exp \left( 9\left( \mu -\frac{1}{2}\Gamma \right) +\frac{21}{2}\left(
\sigma ^{2}+\Gamma \right) \right) \left( \alpha +\beta \right) ^{2}}{%
C4A_{k}^{2}\left( 1-\lambda \right) }  \label{m* Lognormal 1/3 Gamma} \\
&&+\frac{\left( 1-\alpha \right) \delta \lambda \exp \left( 9\left( \mu -%
\frac{1}{2}\Gamma \right) +\frac{15}{2}\left( \sigma ^{2}+\Gamma \right)
\right) \left( \alpha +\beta \right) ^{2}}{C\left( 1-\delta \lambda \right)
4A_{k}^{2}\left( 1-\lambda \right) }  \notag \\
&=&\frac{\left( 1-\alpha \right) \exp \left( 9\mu +\frac{21}{2}\sigma ^{2}+%
\frac{12}{2}\Gamma \right) \left( \alpha +\beta \right) ^{2}}{%
C4A_{k}^{2}\left( 1-\lambda \right) }+\frac{\left( 1-\alpha \right) \delta
\lambda \exp \left( 9\mu +\frac{15}{2}\sigma ^{2}+\frac{6}{2}\Gamma \right)
\left( \alpha +\beta \right) ^{2}}{C\left( 1-\delta \lambda \right)
4A_{k}^{2}\left( 1-\lambda \right) }  \notag
\end{eqnarray}
 }
and it is clear from (\ref{m* lognormal 1/3}) and (\ref{m* Lognormal 1/3
Gamma}) that 
\begin{equation*}
\overline{m}^{\ast }\left( \Gamma \right) -\overline{m}^{\ast }>0
\end{equation*}%
So when $b^{\frac{1}{3}}$ is lognormally distributed we get the result from
proposition \ref{SD} that an increase in the variance that keeps the mean
constant indeed increases $m^{\ast }$. This means that\ a SOSD shift would
reduce $m^{\ast }$.

\section{The roots of the quadratic equation \label{roots}}

Anton et al. (1998) analyze the quadratic case where the function $F(%
\overline{m}_{t-1},\overline{m}_{t},z)$ can be written as 
\begin{equation*}
F(\overline{m}_{t-1},\overline{m}_{t},a,b,c,d,e,f)=\frac{1}{2}a\overline{m}%
_{t-1}^{2}+\frac{1}{2}b\overline{m}_{t}^{2}+c\overline{m}_{_{t-1}}\overline{%
m_{t}}+d\overline{m}_{t-1}+e\overline{m}_{t}+f
\end{equation*}

In this case the roots of the characteristic equation determine the approach
to the steady state. These roots which we denote by $\mu _{i}$ are where $%
\delta $ refers to the discount factor 
\begin{eqnarray*}
\mu _{1} &=&\frac{1}{2}\left( -\frac{b+\delta a}{\delta c}+\sqrt{\left( 
\frac{b+\delta a}{\delta c}\right) ^{2}-\frac{4}{\delta }}\right) \\
\mu _{2} &=&\frac{1}{2}\left( -\frac{b+\delta a}{\delta c}-\sqrt{\left( 
\frac{b+\delta a}{\delta c}\right) ^{2}-\frac{4}{\delta }}\right)
\end{eqnarray*}%
In our case%
\begin{equation*}
a=-C\frac{4A_{k}^{2}\lambda ^{2}}{\left( \left( \alpha +\beta \right) \left(
\int b_{j,t}^{\frac{1}{3}}\right) ^{3}\right) ^{2}}
\end{equation*}%
\begin{equation*}
b=-C\frac{4A_{k}^{2}}{\left( \left( \alpha +\beta \right) \left( \int
b_{j,t}^{\frac{1}{3}}\right) ^{3}\right) ^{2}}
\end{equation*}%
\begin{equation*}
c=C\frac{4A_{k}^{2}\lambda }{\left( \left( \alpha +\beta \right) \left( \int
b_{j,t}^{\frac{1}{3}}\right) ^{3}\right) ^{2}}
\end{equation*}%
and therefore the roots are

\begin{eqnarray*}
\mu _{1} &=&\frac{1}{2}\left( \frac{1+\delta \lambda ^{2}}{\delta \lambda }+%
\sqrt{\left( \frac{1+\delta \lambda ^{2}}{\delta \lambda }\right) ^{2}-\frac{%
4}{\delta }}\right) \\
\mu _{2} &=&\frac{1}{2}\left( \frac{1+\delta \lambda ^{2}}{\delta \lambda }-%
\sqrt{\left( \frac{1+\delta \lambda ^{2}}{\delta \lambda }\right) ^{2}-\frac{%
4}{\delta }}\right)
\end{eqnarray*}

and since $\frac{\partial ^{2}F}{\partial \overline{m}_{t}\partial \overline{%
m}_{t-1}}=C\frac{4A_{k}^{2}\lambda }{\left( \alpha +\beta \right) ^{2}\left(
\int b_{j,t}^{\frac{1}{3}}\right) ^{6}}>0,$ we have that the solution is
determined by $\mu _{2}$. Therefore, 
\begin{equation*}
\overline{m}_{t}=\overline{m}^{\ast }+\mu _{2}^{t}\left( \overline{m}_{0}-%
\overline{m}^{\ast }\right)
\end{equation*}

\end{document}